# *Higher-order symbolic execution for contract verification and refutation*

PHÚC C. NGUYỄN
University of Maryland

SAM TOBIN-HOCHSTADT
Indiana University

DAVID VAN HORN
University of Maryland

## Abstract

We present a new approach to automated reasoning about higher-order programs by endowing symbolic execution with a notion of higher-order, symbolic values.

To validate our approach, we use it to develop and evaluate a system for verifying and refuting behavioral software contracts of components in a functional language, which we call *soft contract verification*. In doing so, we discover a mutually beneficial relation between behavioral contracts and higher-order symbolic execution. Contracts aid symbolic execution by providing a rich language of specifications that can serve as the basis of symbolic higher-order values; the theory of blame enables modular verification and leads to the theorem that *verified components can't be blamed*; and the run-time monitoring of contracts enables *soft* verification whereby verified and unverified components can safely interact and verification is not an all-or-nothing proposition. Conversely, symbolic execution aids contracts by providing compile-time verification which increases assurance and enables optimizations; automated test-case generation for contracts with counter-examples; and engendering a virtuous cycle between verification and the gradual spread of contracts.

Our system uses higher-order symbolic execution, leveraging contracts as a source of symbolic values including unknown behavioral values, and employs an updatable heap of contract invariants to reason about flow-sensitive facts. Whenever a contract is refuted, it reports a concrete *counterexample* reproducing the error, which may involve solving for an unknown function. The approach is able to analyze first-class contracts, recursive data structures, unknown functions, and control-flow-sensitive refinements of values, which are all idiomatic in dynamic languages. It makes effective use of an off-the-shelf solver to decide problems without heavy encodings. Our counterexample search is sound and relatively complete with respect to a first-order solver for base type values. Therefore, it can form the basis of automated verification and bug-finding tools for higher-order programs. The approach is competitive with a wide range of existing tools—including type systems, flow analyzers, and model checkers—on their own benchmarks. We have built a tool which analyzes programs written in Racket, and report on its effectiveness in verifying and refuting contracts.

## Prior publications

This paper unifies and expands upon the work presented in the papers "Soft contract verification," in *Proceedings of the 19th ACM SIGPLAN International Conference on Functional*

*Programming* (Nguyễn et al. 2014) and "Relatively complete counterexamples for higher-order programs," in *Proceedings of the 36th ACM SIGPLAN Conference on Programming Language Design and Implementation* (Nguyễn and Van Horn 2015). It also subsumes the work in the paper "Higher-order symbolic execution via contracts," in *Proceedings of the ACM International Conference on Object Oriented Programming Systems Languages and Applications* (Tobin-Hochstadt and Van Horn 2012).

## 1 Static verification for dynamic languages

Contracts (Meyer 1991; Findler and Felleisen 2002) have become a prominent mechanism for specifying and enforcing invariants in dynamic languages (Disney 2013; Plosch 1997; Austin et al. 2011; Strickland et al. 2012; Hickey et al. 2013). They offer the expressivity and flexibility of programming in a dynamic language, while still giving strong guarantees about the interaction of components. However, there are two downsides: (1) contract monitoring is expensive, often prohibitively so, which causes programmers to write more lax specifications, compromising correctness for efficiency; and (2) contract violations are found only at run-time, which delays discovery of faulty components with the usual negative engineering consequences.

Static verification of contracts would empower programmers to state stronger properties, get immediate feedback on the correctness of their software, and avoid worries about run-time enforcement cost since, once verified, contracts could be removed. All-or-nothing approaches to verification of typed functional programs has seen significant advances in the recent work on static contract checking (Xu et al. 2009; Xu 2012; Vytiniotis et al. 2013), refinement type checking (Terauchi 2010; Zhu and Jagannathan 2013; Vazou et al. 2013, 2014), and model checking (Kobayashi 2009b; Kobayashi et al. 2010, 2011). However, the highly dynamic nature of untyped languages makes verification more difficult.

Programs in dynamic languages are often written in idioms that thwart even simple verification methods such as type inference. Moreover, contracts themselves are written within the host language in the same idiomatic style. This suggests that moving beyond all-or-nothing approaches to verification is necessary.

Fortunately, contracts themselves give us the tools to enable these new approaches, by describing values and by partitioning programs on boundaries. We dub our approach *soft contract verification*, enabling piecemeal and modular verification of contracts. This approach augments a standard reduction semantics for a functional language with contracts and modules by endowing it with a notion of "unknown" values refined by sets of contracts. Verification is carried out by executing programs on abstract values.

Two crucial ideas from contracts allow us to go from whole-program, first-order approaches to modular, higher-order contract verification.

- First, *contracts as abstract values* provide a language of specifications that scales to higher-order values and can encompass arbitrary specifications. This means that whatever guarantees a client needs, they can be specified in the interface and handled by our approach.
- Second, *blame to partition programs* makes modular analysis possible. In a higher-order system, behavioral values can flow across module boundaries. Determining what a modular analysis means in this setting is tricky, but again contracts provide the answer. By re-using the concept of blame from Findler and Felleisen (2002), we define the errors that we rule out as exactly those that blame the portion of the program under consideration. This crucial distinction will become especially important when considering the behavior of unknown higher-order values.

To demonstrate the first step in applying our approach, consider the following contrived, but illustrative example. Let `positive?` and `negative?` be predicates for positive and negative integers. Contracts can be arbitrary predicates, so these functions are also contracts. Consider the following contracted function (written in Racket (Flatt and PLT 2010)):

```
(define/contract (f x)
   (positive? . → . negative?)    ; contract
   (* x -1))
```

We can verify this program by (symbolically) running it on an "unknown" input. Checking the domain contract refines the input to be an unknown satisfying the set of contracts {positive?}. By embedding some basic facts about positive?, negative?, and -1 into the reduction relation for *, we conclude (* {positive?} -1) ⟼ {negative?}, and voilà, we've shown once and for all f meets its contract obligations and cannot be blamed. We could therefore soundly eliminate any contract which blames f, in this case negative?. At its core, we rely on a simple idea: symbolic execution naturally breaks down programs into simpler components, enabling effective reasoning about seemingly-complex features.

This simple approach, building on the two lessons of contracts we have described, is effective for small examples, but insufficient to scale to realistic programs. In this paper, we show how the initial approach can handle tricky problems and larger programs by incorporating several additional techniques.

**Solver-aided reasoning:** While embedding symbolic arithmetic knowledge for specific, known contracts works for simple examples, it fails to reason about arithmetic generally. Contracts often fail to verify because equivalent formulations of contracts are not hard-coded in the semantics of primitives. Many systems address this issue by incorporating an SMT solver. However, for a higher-order language, solver integration is often achieved by reasoning in a theory of uninterpreted functions or semantic embeddings (Knowles and Flanagan 2010; Rondon et al. 2008; Vytiniotis et al. 2013).

In this paper, we observe that higher-order contracts can be effectively verified using only a simple first-order solver. The key insight is that contracts delay higher-order checks and failures always occur with a first order witness. By relying on a (symbolic) semantic approach to carry out higher-order contract monitoring, we can use an SMT solver to reason about integers without the need for sophisticated encodings. (Examples in §2.3.)

**Flow sensitive reasoning:** Just as our semantic approach decomposes higher-order contracts into first-order properties, first-order contracts naturally decompose into conditionals. If the verification procedure did not take this into account, even simple examples would fail to verify:

```
(g : integer? → negative?)
```

```
(define (g x) (if (positive? x) (f x) (f 8)))
```

This is because the true-branch call to `f` is (`f {integer?}`) by substitution, although we know from the guard that `x` satisfies `positive?`.

In this paper, we observe that flow-sensitivity can be achieved by replacing substitution with *heap-allocated* values. These heap addresses are then refined as they flow through predicates and primitive operations, with no need for special handling of contracts (§2.2). As a result, the system is not only effective for contract verification, but can also handle safety verification for programs with no contracts at all.

**First-class contracts:** Pragmatic contract systems enable first-class contracts so new combinators can be written as functions that consume and produce contracts. But to the best of our knowledge, no verification system currently supports first class contracts (or refinements), and in most approaches it appears fundamentally difficult to incorporate such a notion.

Because we handle contracts (and all other features) by *execution*, first-class contracts pose no significant technical challenge and our system reasons about them effectively (§2.6).

**Refuting contracts with concrete counterexamples** Generating inputs that crash first-order programs is a well-studied problem in the literature on symbolic execution (Cadar et al. 2006; Godefroid et al. 2005), type systems (Foster et al. 2002), flow analysis (Xie and Aiken 2005), and software model checking (Yang, Twohey, Engler, and Musuvathi Yang et al.). However, in the setting of higher-order languages, those that treat computations as first-class values, research has largely focused on the verification of programs without investigating how to effectively report counterexamples as concrete inputs when verification fails (e.g., Rondon et al. (2008); Xu et al. (2009); Kawaguchi et al. (2010); Vytiniotis et al. (2013); Tobin-Hochstadt and Van Horn (2012)), or restricted unknown inputs to first-order (e.g., Kobayashi et al. (2011)). Searching for a counterexample witnessing each program bug seems futile in the presence of higher-order unknown inputs: after all, the space of possibilities is huge, and most SMT solvers do not produce models for higher-order unknown values.

Nevertheless, we recognize that even though there are numerous higher-order inputs, they trigger program errors in their contexts following only a few specific patterns. Therefore, instead of searching through the space of all possible functions for a counterexample, we only consider a small subset of functions of specific shapes. The remarkable result is that this method enjoys strong guarantees: each counterexample triggers a real contract violation (*soundness*), and given an SMT solver that is complete for base data types, our method constructs a counterexample reproducing each possible contract violation (*relative completeness*).

**Converging for complex recursion:** Of course, simply executing programs has a fundamental drawback—it will fail to terminate in many cases, and when the inputs are unknown, execution will almost always diverge. Simply detecting cycles in the state space handles straightforward tail-recursive functions, but not more complex recursive calls. Without a solution to this problem, even simple programs operating over inductive data would be impossible to verify.

In this paper, we accelerate the convergence of programs by identifying and approximating regular accumulation of evaluation contexts, causing common recursive programs to converge on unknown values, while providing precise predictions (§2.5). As with the rest of our approach, this happens during execution and is therefore robust to complex, higher-order control flow.

———

Combining these techniques yields a system competitive with a diverse range of existing powerful static checkers, achieving many of their strengths in concert, while balancing the benefits of static contract verification with the flexibility of dynamic enforcement.

We have built a prototype soft verification engine, which we dub SCV, based on these ideas and used it to evaluate the approach (§4). Our evaluation demonstrates that the approach can verify properties typically reserved for approaches that rely on an underlying type system, while simultaneously accommodating the dynamism and idioms of untyped programming languages. We take examples from work on soft typing (Cartwright and Fagan 1991; Wright and Cartwright 1997), type systems for untyped languages (Tobin-Hochstadt and Felleisen 2010), static contract checking (Xu et al. 2009; Xu 2012), refinement type checking (Terauchi 2010), and model checking of higher-order functional languages (Kobayashi 2009b; Kobayashi et al. 2010, 2011).

SCV can prove all contract checks redundant for almost all of the examples taken from this broad array of existing program analysis and type checking work, and can handle many of the tricky higher-order verification problems demonstrated by other systems. In other words, our approach is competitive with type systems, model checkers, and soft typing systems on each of their chosen benchmarks—in contrast, work on higher-order model checking does not handle benchmarks aimed at soft typing or occurrence typing, and vice versa. In the cases where SCV does not prove the complete absence of contract errors, the vast majority of possible dynamic errors are ruled out, justifying numerous potential optimizations. Over this corpus of programs, 99% of the contract and run-time type checks are proved safe, and could be eliminated.

We also evaluate the verification of three small interactive video games which use first-class and dependent contracts pervasively. The results show the subsequent elimination of contract monitoring has a dramatic effect: from a speed up factor of 7 in one case, to three orders of magnitude in the others. In essence, these results show the games are infeasible without contract verification.

## 2 Worked examples

We now present the main ideas of our approach through a series of examples taken from work on other verification techniques, starting from the simplest and working up to a complex object encoding.

### *2.1 Higher-order symbolic reasoning*

Consider the following simple function that transforms functions on even integers into functions on odd integers. It has been ascribed this specification as a contract, which can be monitored at run-time.

```
(e2o : (even? → even?) → (odd? → odd?))
(define (e2o f)
  (λ (n) (- (f (+ n 1)) 1)))
```

A contract monitors the flow of values between components. In this case, the contract monitors the interaction between the context and the e2o function. It is easy to confirm that e2o is correct with respect to the contract; e2o holds up its end of the agreement, and therefore cannot be blamed for any run-time failures that may arise. The informal reasoning goes like this: First assume f is an even? → even? function. When applied, we must ensure the argument is even (otherwise e2o is at fault), but may assume the result is even (otherwise the context is at fault). Next assume n is odd (otherwise the context is at fault) and ensure the result is odd (otherwise e2o is at fault). Since (+ n 1) is even when n is odd, f is applied to an even argument, producing an even result. Subtracting one therefore gives an odd result, as desired.

This kind of reasoning mimics the step-by-step computation of e2o, but rather than considering some particular inputs, it considers these inputs symbolically to verify all possible executions of e2o. We systematize this kind of reasoning by augmenting a standard reduction semantics for contracts with symbolic values that are refined by sets of contracts. At first approximation, the semantics includes reductions such as:

$$(+ \ \{\texttt{odd?}\} \ 1) \longmapsto \{\texttt{even?}\}, \text{ and}$$
$$(\{\texttt{even?} \rightarrow \texttt{even?}\} \ \{\texttt{even?}\}) \longmapsto \{\texttt{even?}\}.$$

This kind of symbolic reasoning mimics a programmer's informal intuitions which employ contracts to refine unknown values and to verify components meet their specifications. If a component cannot be blamed in the symbolic semantics, we can safely conclude it cannot be blamed in general.

### *2.2 Flow sensitive reasoning*

Programmers using untyped languages often use a mixture of type-based and flow-based reasoning to design programs. The analysis naturally takes advantage of type tests idiomatic in dynamic languages even when the tests are buried in complex expressions. The following function taken from work on occurrence typing (Tobin-Hochstadt and Felleisen 2010) can be proven safe using our symbolic semantics:

```
(f : (or/c int? str?) cons? → int?)
(define (f x p)
  (cond
    [(and (int? x) (int? (car p))) (+ x (car p))]
    [(int? (car p)) (+ (str-len x) (car p))]
    [else 0]))
```

Here, `int?`, `str?`, and `cons?` are type predicates for integers, strings, and pairs, respectively. The contract (`or/c int? str?`) uses the `or/c` contract combinator to construct a contract specifying a value is either an integer or a string.

A programmer would convince themselves this program was safe by using the control dominating predicates to refine the types of x and (car p) in each branch of the conditional.[1] Our symbolic semantics accommodates exactly this kind of reasoning in order to verify this example. However, there is a technical challenge here. A straightforward substitution-based semantics would not reflect the flow-sensitive facts. Focusing just on the first clause, a substitution model would give:

```
(cond
  [(and (int? {(or/c int? str?)}) (int? (car {cons?})))
   (+ {(or/c int? str?)} (car {cons?}))] …)
```

At this point, it's too late to communicate the refinement of these sets implied by the test evaluating to true, so the semantics would report the contract on + potentially being violated because the first argument may be a string, and the second argument may be anything. We overcome this challenge by modelling symbolic values as heap-allocated sets of contracts. When predicates and data structure accessors are applied to heap addresses, we refine the corresponding sets to reflect what must be true. So the program is modelled as:

```
(cond
  [(and (int? L₁) (int? (car L₂)))
   (+ L₁ (car L₂))] …)
```
where $L_1 \mapsto \{\texttt{(or/c int? string?)}\}$, $L_2 \mapsto \{\texttt{cons?}\}$

In the course of evaluating the test, we get to (`int?` $L_1$), the semantics conceptually forks the evaluator and refines the heap:

$$(\texttt{int?} \ L_1) \longmapsto \texttt{true}, \text{ where } L_1 \mapsto \{\texttt{int?}\}$$
$$\longmapsto \texttt{false}, \text{ where } L_1 \mapsto \{\texttt{string?}\}$$

Similar refinements to $L_2$ are communicated through the heap for (`int?` (`car` $L_2$)), thereby making (+ $L_1$ (`car` $L_2$)) safe. This simple idea is effective in achieving flow-based refinements. It naturally handles deeply nested and inter-procedural conditionals.

### 2.3 Incorporating an SMT solver

The techniques described so far are highly effective for reasoning about functions and many kinds of recursive data structures. However, effective reasoning about many kinds of base values, such as integers, requires sophisticated domain-specific knowledge. Rather than build such a tool ourselves, we defer to existing high-quality solvers for these domains. Unlike many solver-aided verification tools, however, we use the solver *only* for queries on base values, rather than attempting to encode a rich, higher-order language into one that is accepted by the solver. This obviates the need of a general (and error-prone) translation of the language. For example, there is no need to embed an untyped language's "unityped type system" into the solver's type system.

---

[1] The call to `str-len` is safe because (`and` (`int?` x) (`int?` (`car` p))) being false and (`int?` (`car` p)) being true implies that (`int?` x) is false, which in turns implies x is a string as enforced by f's contract.

To demonstrate our approach, we take an example (`intro3`) from work on model checking higher-order programs (Kobayashi et al. 2011).

```
; (>/c n) abbreviates (λ (x) (> x n))

(define (f x g) (g (+ x 1)))

(h : [x : int?] → [y : (and/c int? (>/c x))] → (and/c int? (>/c y)))
(define (h x) ...) ; unknown definition

(main : int? → (and/c int? (>/c 0)))
(define (main n) (if (≥ n 0) (f n (h n)) 1))
```

In this program, we define a contract combinator (`>/c`) that creates a check for an integer from a lower bound; a helper function `f`, which comes without a contract; and an *unknown* function `h` that given an integer `x`, returns a function mapping some number `y` that is greater than `x` to an answer greater than `y`—here `h`'s specification is given, but not its implementation. (Note `h`'s contract is dependent.) We verify `main`'s correctness, which means it definitely returns a positive integer and does not violate `h`'s contract.

According to its contract, `main` is passed an integer `n`. If `n` is negative, `main` returns `1`, satisfying the contract. Otherwise the function applies `f` to `n` and `(h n)`. Function `h`, by its contract, returns another function that requires a number greater than `n`. Examining `f`'s definition, we see `h` (now bound to `g`) is eventually applied to `(+ n 1)`. Let $n_1$ be the result of `(+ n 1)`. And by `h`'s contract, we know the answer to `(h n)` is another integer greater than $n_1$. Let us name this answer $n_2$. In order to verify that `main` satisfies contract `(>/c 0)`, we need to verify that $n_2$ is a positive integer.

Once `f` returns, the heap contains several addresses with contracts:

$$\begin{aligned} n &\mapsto \{\texttt{int?}, (\geq/\texttt{c 0})\} \\ n_1 &\mapsto \{\texttt{int?}, (=/\texttt{c (+ n 1)})\} \\ n_2 &\mapsto \{\texttt{int?}, (>/\texttt{c } n_1)\} \end{aligned}$$

We then translate this information to a query for an external solver:

```
n, n₁, n₂: INT;
ASSERT n ≥ 0;
ASSERT n₁ = n + 1;
ASSERT n₂ > n₁;
QUERY n₂ > 0;
```

Solvers such as CVC4 (Barrett et al. 2011) and Z3 (Moura and Bjørner 2008) easily verify this implication, proving `main`'s correctness.

Refinements such as (≥/c 0) are generated by *primitive* applications (≥ x 0), and queries are generated from translation of the heap, not arbitrary expressions. This has a few consequences. First, by the time we have value `v` satisfying predicate `p` on the heap, we know that `p` terminates successfully on `v`. Issues such as errors (from `p` itself) or divergence are handled elsewhere in other evaluation branches. Second, we only need to translate

*Higher-order symbolic execution for contract verification and refutation*     9

a small set of simple, well understood contracts—not arbitrary expressions. Evaluation naturally breaks down complex expressions, and properties are discovered even when they are buried in complex, higher-order functions. Given a translation for (>/c 0), the analysis automatically takes advantage of the solver even when the predicate contains > in a complex way, such as (λ (x) (or (> x 0) *E*) where *E* is an arbitrary expression. Predicates that lack translations to SMT only reduce precision, never soundness.

### *2.4 Generating higher-order counterexamples*

Programmers benefit not only from verification of correct programs, but also refutation of incorrect programs through concrete counterexamples: they minimize the confusion between a true bug and a false warning, and provide programmers with insight into their code's defects.

In the following program, f's contract promises that if its argument is a function returning an integer, then f returns an integer. In its body, f performs a division involving the application of its argument to 42.

```
(f : (int? → int?) → int?)
(define (f g)
  (/ 1 (- 100 (g 42))))
```

Function f's definition is unsafe in two ways. First, the division is not protected against a denumerator of 0. Second, / potentially returns a quotient, causing f to violate the int? contract in its range.

In this case, the only way function g interacts with the code under verification (function f) is through its returned value. Because g is applied only to 42 in this case, it suffices to search for instantiations of g in the space of constant functions of the form (λ (_) n) with n being an unknown integer. The system therefore can produce two counterexaples trigger two potential bugs in the program:

```
Contract violation: f violates contract with /
Value 0 violates contract (not/c (=/c 0))
An example that triggers this violation:
    (f (λ (n) 100))

Contract violation: f violates its own contract
Value -1/2 violates contract int?
An example that triggers this violation:
    (f (λ (n) 102))
```

In more complex programs, an unknown function can interact and trigger errors in multiple ways: either by returning another value to be consumed by the context, or by applying a function coming from the context to some values. The counterexample can be more complex, but in each case, it is only how the unknown function interacts with its context that is relevant to producing a counterexample. The system needs not consider instantiations to unknown functions that perform irrelevant work, diverge, or have their own errors.

### *2.5 Converging for non-tail recursion*

The techniques sketched above provide high precision in the examples considered, but simply executing programs on abstract values is unlikely to terminate in the presence of recursion. When an abstract value stands for an infinite set of concrete values, execution may unfold infinitely, building up an ever-growing evaluation context. To tackle this problem, we *summarize* this context to coalesce repeated structures and enable termination on many recursive programs. Although guaranteed termination is not our goal, the empirical results (§4) demonstrate that the method is effective in practice.

The following example program is taken from work on model checking of higher-order functional programs (Kobayashi et al. 2011), and demonstrates checking non-trivial safety properties on recursive functions. Note that no loop invariants need be provided by the user.

```
(main : (and/c int? (>=/c 0)) → (and/c int? (>=/c 0)))
(define (main n)
  (let ([l (make-list n)])
    (if (> n 0) (car (reverse l empty)) 0)))

(define (reverse l ac)
  (if (empty? l) ac
      (reverse (cdr l) (cons (car l) ac))))

(define (make-list n)
  (if (= n 0) empty
      (cons n (make-list (- n 1)))))
```

Again, we aim to verify both the specified contract for main as well as the preconditions for primitive operations such as car. Most significantly, we need to verify that (reverse l empty) produces a non-empty list (so that car succeeds) and that its first element is a positive integer. The local functions reverse and make-list do not come with a contract.

This problem is more challenging than the original OCaml version of the same program, due to the lack of types. This program represents a common idiom in dynamic languages: not all values are contracted, and there is no type system on which to piggy-back verification. In addition, programmers often rely on inter-procedural reasoning to justify their code's correctness, as here with reverse.

We verify main by applying it to an abstract (unknown) value $n_1$. The contract ensures that within the body, $n_1$ is a non-negative integer.

The integer $n_1$ is first passed to make-list. The comparison (= $n_1$ 0) non-deterministically returns true or false, updating the information known about $n_1$ to be either 0 or (>/c 0) in each corresponding case. In the first case, make-list returns empty. In the second case, make-list proceeds to the recursive application (make-list $n_2$), where $n_2$ is the abstract non-negative integer obtained from evaluating (- $n_1$ 1). However, (make-list $n_2$) is identical to the original call (make-list $n_1$) up to renaming, since both $n_1$ and $n_2$ are non-negative. Therefore, we pause here and use a summary of make-list's result instead of continuing in an infinite loop.

Since we already know that empty is one possible result of (make-list $n_1$), we use it as the result of (make-list $n_2$). The application (make-list $n_1$) therefore produces the pair ⟨n₁,empty⟩, which is another answer for the original application. We could continue this process and plug this new result into the pending application (make-list n₂). But we instead first approximate ⟨n₁,empty⟩ to a non-empty list of positive integers. This approximation choice is guided by the observation that plugging in empty in the recursive call gives rise to ⟨n₁,empty⟩. We then use this approximate answer as the result of the pending application (make-list n₂). This then induces another result for (make-list n₁), a list of two or more positive integers, but this is subsumed by the previous answer of non-empty integer list. We have now discovered *all* possible return values of make-list when applied to a non-negative integer: it maps 0 to empty, and positive integers to a non-empty list of positive integers.

Although our explanation made use of the order, the soundness of analyzing make-list does not depend on the order of exploring non-deterministic branches. Each recursive application with repeated arguments generates a waiting context, and each function return generates a new case to resume. There is an implicit work-list algorithm in the modified semantics (§3.8.2).

When make-list returns to main, we have two separate cases: either n₁ is 0 and l is empty, or n₁ is positive and l is non-empty. In the first case, (> n₁ 0) is false and main returns 0, satisfying the contract. Otherwise, main proceeds to reversing the list before taking its first element.

Using the same mechanism as with make-list, the analysis infers that reverse returns a non-empty list when either of its arguments (l or acc) is non-empty. In addition, reverse only receives arguments of proper lists, so all partial operations on l such as car and cdr are safe when l is not empty, without needing an explicit check. The function eventually returns a non-empty list of integers to main, justifying main's call to the partial function car, producing a positive integer. Thus, main never has a run-time error in any context.

While this analysis makes use of the implementation of make-list and reverse, that does not imply that it is whole-program. Instead, it is *modular* in its use of unknown values abstracting arbitrary behavior. For example, make-list could instead be an abstract value represented by a contract that always produces lists of integers. The analysis would still succeed in proving all contracts safe except the use of car in main—this shows the flexibility available in choosing between precision and modularity. In addition, the analysis does not have to be perfectly precise to be useful. If it successfully verifies most contracts in a module, that already greatly improves confidence about the module's correctness and justifies the elimination of numerous expensive dynamic checks.

### 2.6 First-class contracts

In the following, we choose a simple encoding of classes as functions that produce objects, where objects are again functions that respond to messages named by symbols. We then verify the correctness of a *mixin*: a function from classes to classes. The vec/c contract enforces the interface of a 2D-vector class whose objects accept messages 'x, 'y, and 'add for extracting components and vector addition.

```
(define vec/c
  ([msg : (one-of/c 'x 'y 'add)]
   → (match msg
```

```
            [(or 'x 'y) real?]
            ['add (vec/c → vec/c)]])))
```

This definition demonstrates several powerful contract system features which we are able to handle:

- contracts are first-class values, as in the definition of vec/c,
- contracts may include arbitrary predicates, such as real?,
- contracts may be recursive, as in the contract for 'add,
- function contracts may express *dependent* relationships between the domain and range—the contract of the result of method selection for vec/c depends on which method is chosen.

Suppose we want to define a mixin that takes any class that satisfies the vec/c interface and produces another class with added vector operations such as 'len for computing the vector's length. The extend function defines this mixin, and ext-vec/c specifies the new interface. We verify that extend violates no contracts and returns a class that respects specifications from ext-vec/c.

```
(extend : (real? real? → vec/c) → (real? real? → ext-vec/c))
(define (extend mk-vec)
  (λ (x y)
    (let ([vec (mk-vec x y)])
      (λ (m)
        (match m
          ['len
           (let ([x (vec 'x)] [y (vec 'y)])
             (sqrt (+ (* x x) (* y y))))]
          [_ (vec m)])))))

(define ext-vec/c
  ([msg : (one-of/c 'x 'y 'add 'len)]
   → (match msg
       [(or 'x 'y) real?]
       ['add (vec/c → vec/c)]
       ['len (and/c real? (≥/c 0))])))
```

To verify extend, we provide an arbitrary value, which is guaranteed by its contract to be a class matching vec/c. The mixin returns a new class whose objects understand messages 'x, 'y, 'add, and 'len. This new class defines method 'len and relies on the underlying class to respond to 'x, 'y, and 'add. Because the old class is constrained by contract vec/c, the new class will not violate its contract when responding to messages 'x, 'y, and 'add.

For the 'len message, the object in the new vector class extracts its components as abstract numbers x and y, according to interface vec/c. It then computes their squares and leaves the following information on the heap:

$$\begin{aligned} x^2 &\mapsto \{\text{real?},(\text{=/c } (* \text{ x x}))\} \\ y^2 &\mapsto \{\text{real?},(\text{=/c } (* \text{ y y}))\} \\ s &\mapsto \{\text{real?},(\text{=/c } (+ \text{ x}^2 \text{ y}^2))\} \end{aligned}$$

Solvers such as Z3 (Moura and Bjørner 2008) can handle simple non-linear arithmetic and verify that the sum `s` is non-negative, thus the `sqrt` operation is safe. Execution proceeds to take the square root—now called `l`—and refines the heap with the following mapping:

$$l \mapsto \{\mathtt{real?},(\mathtt{=/c\ (sqrt\ s)})\}$$

When the method returns, its result is checked by contract `ext-vec/c` to be a non-negative number. We again rely on the solver to prove that this is the case.

Therefore, `extend` is guaranteed to produce a new class that is correct with respect to interface `vec-ext/c`, justifying the elimination of expensive run-time checks. In a Racket program computing the length of 100000 random vectors, eliminating these contracts results in a 100-fold speed-up. While such dramatic results are unlikely in full programs, measurements of existing Racket programs suggests that 50% speedups are possible (Strickland et al. 2012).

## 3 A Symbolic Language with Contracts

In this section, we give a reduction system describing the core of our approach. Symbolic $\lambda_C$ is a model of a pure functional language with first-class contracts and *symbolic values*. We first present the semantics, including handling of primitives and unknown functions, that facilitates finding bugs and constructing test cases reproducing each reachable contract violation. We then describe how the handling of primitive values integrates with external solvers. Finally, we show an abstraction of our system to accelerate convergence, turning the bug-finding semantics into a practical verification. For each abstraction, we relate concrete and symbolic programs and prove a soundness theorem.

At a high level, the key idea of our semantics is that abstract values behave non-deterministically in all possible ways that concrete values might behave. Furthermore, abstract values can be bounded by specifications in the form of contracts that limit these behaviors. As a result, an operational semantics for abstract values explores all the ways that the concrete program under consideration might be used.

Given this operational semantics, we can then examine the results of evaluation to see if any results are errors blaming the components we wish to verify. If they do not, then our soundness theorem implies that there are no ways for the component to be blamed, regardless of what other parts of the program do. Thus, we have verified the component against its contract in all contexts. We make this notion precise in section 3.6.

### 3.1 Syntax of Symbolic $\lambda_C$

Our initial language models the functional core of many modern dynamic languages, extended with behavioral, first-class contracts, as well as symbolic values. The abstract syntax is shown in figure 1. Syntax needed only for symbolic execution is highlighted in gray; we discuss it after the syntax of concrete programs.

A program *P* is a sequence of module definitions followed by a top-level expression which may reference the modules. Each module *M* has a name *H* and exports a single value *V* with behavior enforced by contract $V_c$. (Generalization to multiple-export modules is straightforward.)

| | | | |
|---|---|---|---|
| Programs | $P,Q$ | ::= | $\vec{M}\,E$ |
| Modules | $M$ | ::= | $(\text{module}\,H\,V_c\,V)$ |
| Expressions | $E,C$ | ::= | $A \mid X^H \mid E\,E^H \mid O\,\vec{E}^H \mid \text{if}\,E\,E\,E \mid C \rightarrow \lambda X.C$ |
| | | $\mid$ | $\text{mon}_H^{H,H}(C,E) \mid \texttt{assume}(V,V)$ |
| Answers | $A$ | ::= | $V \mid \texttt{blame}_H^H$ |
| Values | $V$ | ::= | $U \mid L$ |
| Concrete Values | $U$ | ::= | $n \mid \lambda X.E \mid V \rightarrow \lambda X.C$ |
| Operations | $O$ | ::= | $O_? \mid \texttt{add1} \mid + \mid = \mid \ldots$ |
| Predicates | $O_?$ | ::= | $\texttt{zero?} \mid \texttt{int?} \mid \texttt{proc?} \mid \texttt{dep?}$ |
| Variables | $X, H$ | $\in$ | *identifier* |
| Addresses | $L$ | $\in$ | *address* |

Fig. 1. Syntax of Symbolic $\lambda_C$

Expressions include standard forms such as values, variable and module references, applications, and conditionals, as well as those for constructing and monitoring contracts. Contracts are first-class values and can be produced by arbitrary expressions. For clarity, when an expression plays the role of a contract, we use the metavariable $C$, rather than $E$. A *dependent* function contract ($C \rightarrow \lambda X.C'$) monitors a function's argument with $C$ and its result with the contract produced by applying $\lambda X.C'$ to the argument.

A contract violation at run-time causes *blame*, an error with information about who violated the contract. We write $\texttt{blame}_{H''}^H$ to mean module $H$ is blamed for violating the contract from $H''$. The form ($\text{mon}_{H''}^{H,H'}(C,E)$) monitors expression $E$ with contract $C$, with $H$ being the positive party, $H'$ the negative party, and $H''$ the source of the contract. The system blames the positive party if $E$ produces a value violating $C$, and the negative one if $E$ is misused by the *context* of the contract check. To make context information available at run-time, we annotate references and applications with labels indicating the module they appear in, or † for the top-level expression. For example, $H^{H'}$ denotes a reference to the name $H$ from module $H'$, and $(\texttt{add1}\,X)^\dagger$ denotes an addition inside the top level. When a module $H$ causes a primitive error, such as applying 5, we also write $\texttt{blame}_\Lambda^H$, indicating that it violates a contract with the language. Monitoring forms, blaming forms, and labels are not available for programmers to write. We omit labels when they are irrelevant or can be inferred.

*Concrete values U* include abstractions, integers, and dependent contracts with domain components evaluated. We use 0 to indicate falsehood and any other value for truth. Primitive operations over values are standard, including predicates $O_?$ for dynamic testing of data types.

To reason about absent components, we equip $\lambda_C$ with *unknown*, or *symbolic values*, which abstract over multiple concrete values exhibiting a range of behavior. Each address $L$ identifies an arbitrary but fixed and syntactically closed[2] value in the program. For soundness, execution must account for *all* possible concretizations of unknown values, and reduction becomes non-deterministic. As execution progresses through tests and contract checks, more assumptions can be made about symbolic values in each non-deterministic branch. To track refinements of symbolic values, we use a heap that maps each address to a refinable value, which includes concrete as well as abstract values of the form $\bullet^{\overrightarrow{U}}$ and $\mathsf{case}[\overrightarrow{V \mapsto L}]$. We omit displaying this predicate set when it is empty, irrelevant, or can be inferred from context. The form $\bullet^{\overrightarrow{U}}$ denotes a value known to satisfy contract set $\overrightarrow{U}$ but is otherwise unknown. The form $\mathsf{case}[\overrightarrow{V \mapsto L}]$ is used internally and denotes a mapping between values, which we discuss further in section 3.2.3.

---

[2] For example, $L$ cannot be instantiated by term ($\lambda$x.y)

$$\begin{aligned}
\text{Refined Values } U' &::= U \mid \bullet^{\overrightarrow{U}} \mid \mathsf{case}[\overrightarrow{V \mapsto L}] \\
\text{Heaps} \qquad \Sigma &::= \overrightarrow{\langle L, U' \rangle}.
\end{aligned}$$

### 3.2 Semantics of Symbolic $\lambda_C$

We now turn to the reduction semantics for Symbolic $\lambda_C$, which combines standard rules for untyped languages with behavior for unknown values. Reduction is defined as a relation on states, parameterized by a module context. We omit the module context whenever it is irrelevant.

$$\overrightarrow{M} \vdash \varsigma \longmapsto \varsigma'$$

A state is an expression paired with a heap:

$$\text{States} \quad \varsigma ::= \langle E, \Sigma \rangle.$$

#### 3.2.1 Basic rules

The first reduction rule concerns the application of primitive operations, which are interpreted by a $\delta$ relation. The relation maps operations, arguments and heaps to results and new heaps.

$$\frac{\delta(\Sigma, O, \overrightarrow{V}) \ni \varsigma}{(O\ \overrightarrow{V}), \Sigma \longmapsto \varsigma} \textit{Apply-Primitive}$$

The use of a $\delta$ relation in reduction semantics is standard, but typically it is a function and is independent of the heap. We make $\delta$ dependent on the heap in order to use and update the current set of invariants; we make it a relation, since it may behave non-deterministically on unknown values. For example, in interpreting (> L 5) where $L \mapsto \bullet^{\texttt{int?}}$, the $\delta$ relation will produce two results: 1, with an updated heap to reflect the unknown value is (>/c 5); the other 0, with a heap reflecting the opposite. The $\delta$ relation is thus the hub of the verification system and a point of interaction with the SMT solver. It is described in more detail in section 3.3.

Applications of $\lambda$-abstractions follow standard $\beta$-reduction; applications of non-functions result in blame.

$$\textit{Apply-Function} \qquad\qquad \textit{Apply-Non-Function}$$

$$\frac{}{(\lambda X.E\ V), \Sigma \longmapsto [V/X]E, \Sigma} \qquad \frac{\delta(\Sigma, \mathsf{proc?}, V) \ni \langle 0, \Sigma' \rangle}{(V\ V')^H, \Sigma \longmapsto \mathsf{blame}^H_\Lambda, \Sigma'}$$

Notice that in rule *Apply-Non-Function*, the $\delta$ relation is employed to determine whether the value in operator position is a function using the `proc?` primitive. (Non-functions include concrete numbers as well as abstract values known to exclude functions; application of abstract values that may be functions is described in section 3.2.3.)

Conditionals treat values other than `0` as true.

$$\frac{\textit{If-True}}{\delta(\Sigma,\texttt{zero?},V) \ni \langle 0, \Sigma' \rangle} \qquad \frac{\textit{If-False}}{\delta(\Sigma,\texttt{zero?},V) \ni \langle 1, \Sigma' \rangle}$$
$$\textsf{if } V\, E_1\, E_2, \Sigma \longmapsto E_1, \Sigma' \qquad \textsf{if } V\, E_1\, E_2, \Sigma \longmapsto E_2, \Sigma'$$

Just as in the case of *Apply-Non-Function*, the interpretation of conditionals uses the $\delta$ relation to determine whether `zero?` holds, which takes into account all of the knowledge accumulated in $\Sigma$ and in either branch that is taken, updates the current knowledge to reflect whether `zero?` of $V$ holds. This is the mechanism by which control-flow based refinements are enabled.

The two rules for module references reflect the approach in which contracts are treated as *boundaries* between components (Dimoulas et al. 2011): a module self-reference incurs no contract check, while cross-module references are protected by the specified contract.

$$\frac{\textit{Module-Self-Reference}}{(\texttt{module}\, H\, V_c\, V) \in \overrightarrow{M}} \qquad \frac{\textit{Module-External-Reference}}{(\texttt{module}\, H\, V_c\, V) \in \overrightarrow{M} \quad H \neq H'}$$
$$\overrightarrow{M} \vdash H^H, \Sigma \longmapsto V, \Sigma \qquad \overrightarrow{M} \vdash H^{H'}, \Sigma \longmapsto \textsf{mon}_H^{H,H'}(V_c, V), \Sigma$$

Finally, any state that is stuck with blame inside an evaluation context transitions to a final blame state that discards the surrounding context.

$$\frac{\textit{Halt-Blame}}{\mathscr{E} \neq [\,]}$$
$$\mathscr{E}[\texttt{blame}], \Sigma \longmapsto \texttt{blame}, \Sigma$$

Evaluation contexts are defined as follows:

$$\mathscr{E} ::= \quad [\,] \mid \mathscr{E}\, E \mid V\, \mathscr{E} \mid O\, \overrightarrow{V}\, \mathscr{E}\, \overrightarrow{E} \mid \textsf{if}\, \mathscr{E}\, E\, E \mid \textsf{mon}(\mathscr{E}, E) \mid \textsf{mon}(V, \mathscr{E}) \mid \mathscr{E} \rightarrow \lambda X.E$$

### 3.2.2 Contract monitoring

Contract monitoring follows existing operational semantics for contracts (Findler and Felleisen 2002), with extensions to handle and refine symbolic values.

There are several cases for checking a value against a contract. If the contract is not a function contract, we say it is *flat*, denoting a first-order property to be checked immediately. We thus expand the checking expression to a conditional.

$$\frac{\textit{Monitor-Flat-Contract}}{\delta(\Sigma, \texttt{dep?}, V_c) \ni \langle 0, \Sigma' \rangle \qquad \Sigma' \vdash V : V_c\, \textbf{?}}$$
$$\textsf{mon}_{H''}^{H,H'}(V_c, V), \Sigma \longmapsto \textsf{if}\, (V_c\, V)\, \texttt{assume}(V, V_c)\, \texttt{blame}_{H''}^H, \Sigma'$$

Since contracts are first-class, they can also be abstract values; we rely on $\delta$ to determine whether a value is a flat contract by using (the negation of) the predicate for dependent contracts, `dep?`, instead of examining the syntax. This rule is standard except for the use of $\texttt{assume}(V, V_c)$ and the $(\cdot \vdash \cdot : \cdot\, \textbf{?})$ judgment. The $\texttt{assume}(V, V_c)$ form, which would normally

*Higher-order symbolic execution for contract verification and refutation* 17

just be $V$, dynamically refines address $V$ in the heap to indicate that $V$ satisfies $V_c$; assume is discussed further in section 3.2.3. The judgment $\Sigma' \vdash L : V$ **?**, which would normally just be omitted, indicates that the contract $V$ cannot be statically judged to either pass or fail for $L$, which is why the predicate must be applied. This judgment and its closely related counterparts $(\cdot \vdash \cdot : \cdot \checkmark)$ and $(\cdot \vdash \cdot : \cdot \text{✗})$, which statically proves a value must or must not satisfy a given contract respectively, are discussed in section 3.4.

If a flat contract can be statically proved or refuted, monitoring can be short-circuited.

*Monitor-Proved*
$$\frac{\delta(\Sigma,\text{dep?},V_c) \ni \langle 0, \Sigma' \rangle \quad \Sigma' \vdash V : V_c \checkmark}{\text{mon}(V_c,V),\Sigma \longmapsto V,\Sigma'}$$

*Monitor-Refuted*
$$\frac{\delta(\Sigma,\text{dep?},V_c) \ni \langle 0, \Sigma' \rangle \quad \Sigma' \vdash V : V_c \text{✗}}{\text{mon}(V_c,V),\Sigma \longmapsto \texttt{blame},\Sigma'}$$

Monitoring a function contract against a function is interpreted the standard $\eta$-expansion of contracts, where we swap the blame roles of positive and negative parties (Findler and Felleisen 2002). Similar to other values, function contracts can be either concrete or symbolic. As we later shown in the definitions of $\delta$ and helper metafunction *refine*, when a symbolic value is assumed a dependent contract, we decompose it into 2 other symbolic values identifying its domain and range.

*Monitor-Function-Contract*
$$\frac{\delta(\Sigma,\text{proc?},V) \ni \langle 1, \Sigma' \rangle}{\text{mon}_{H''}^{H,H'}(V_c \rightarrow \lambda X.C,V),\Sigma \longmapsto \lambda X.\text{mon}_{H''}^{H,H'}(C,(V \; \text{mon}_{H''}^{H',H}(V_c,X))),\Sigma'}$$

*Monitor-Abstract-Function-Contract*
$$\frac{\delta(\Sigma,\text{proc?},V) \ni \langle 1, \Sigma' \rangle \quad \delta(\Sigma',\text{dep?},L) \ni \langle 1, \Sigma'' \rangle \quad \Sigma''(L) = V_c \rightarrow \lambda X.C}{\text{mon}(L,V),\Sigma \longmapsto \lambda X.\text{mon}_{H''}^{H,H'}(C,(V \; \text{mon}_{H''}^{H',H}(V_c,X))),\Sigma''}$$

Finally, monitoring a function contract against a non-function results in an error blaming the party providing the value.

*Monitor-Non-Function*
$$\frac{\delta(\Sigma,\text{dep?},V_c) \ni \langle 1, \Sigma_1 \rangle \quad \delta(\Sigma_1,\text{proc?},V) \ni \langle 0, \Sigma_2 \rangle}{\text{mon}_{H''}^{H,H'}(V_c,V),\Sigma \longmapsto \texttt{blame}_{H''}^{H},\Sigma_2}$$

### 3.2.3 Handling unknown values

The assume form updates the heap of refinements to take into account the new information using the refine metafunction; see figure 2 for the definition of *refine*.

*Assume*
$$\overline{\texttt{assume}(V,V_c),\Sigma \longmapsto V, \textit{refine}(\Sigma,V,V_c)}$$

Refinement is straightforward propagation of known contracts, expanding values known to be function contracts (via dep?) into function contract values.

Finally, we must handle application of unknown values. Notice that in the presence of higher-order arguments, the obvious solution of using a table to model each unknown function does not work. First, a higher-order function interacts with its context not only through its returned result, but also the values it supplies to its functional arguments. Using a table would omit the latter means through which the unknown function triggers errors in its context. Second, there is no obvious choice of equality between higher-order values for use as table keys. For example, it is not clear whether (λ (x) x) and (λ (y) y) should be the same key. Third, there is no direct connection between a higher-order keyed table and a λ-term: a naive construction does not yield the intended function. The following λ-term indeed would only execute the else clause reguardless of its argument, because a comparison to a function literal in most languages is guaranteed to return false.

```
(λ (f)
 (cond [(equal? f (λ (x) x)) ...#|dead code|#...]
       [(equal? f (λ (y) y)) ...#|dead code|#...]
       [else ...]))
```

Therefore, instead of viewing a higher-order function as a mapping, we consider different ways in which it interacts with the known components of the program. Even though there are numerous ways to instantiate a λ-term, a function only interacts with its context in a few specific ways. For example, it is not neccessary to consider unknown components that perform unnecessary computations, have their own errors, or diverge: for each function with such behavior, there exists another terminating, error-free function that explores no fewer contract violations in concrete modules.

We therefore refine each unknown function to have a specific shape shown in rule *Apply-Unknown*. The unknown function dynamically inspects its argument's datatype to perform an appropriate operation. If the argument is a first-order value, we model the function as a table using the $\mathsf{case}[\overrightarrow{V \mapsto L}]$ form discussed next. If the argument is a function, the unknown function can interact with it in several ways: (1) apply the function to an unknown value then pass the result to another unknown "continuation", (2) delay the exploration of the function's behavior and return a value depending on this function, (3) ignore the function and return a value independent of it. We use addresses $L_f, L_g, L_x, L_a$ for new symbolic values that this unknown function decomposes to, and symbolic values $L_1, L_2$ to encode the non-deterministic (but remembered) choices.

*Apply-Unknown*
$$\frac{\delta(\Sigma, \mathsf{proc?}, L) \ni \langle 1, \Sigma_1 \rangle \qquad \Sigma_1(L) = \bullet \\ \Sigma_2 = \Sigma_1[L_1 \mapsto \bullet, L_2 \mapsto \bullet, L_f \mapsto \bullet, L_g \mapsto \bullet, L_x \mapsto \bullet, L_a \mapsto \bullet, L' \mapsto \mathsf{case}[\,], L \mapsto \lambda X.E] \\ \text{where } E = \mathsf{if}\,(\mathsf{proc?}\,X)\,(\mathsf{if}\,L_1\,((L_f\,(X\,L_x))\,X)\,(\mathsf{if}\,L_2\,\lambda Y.((L_g\,X)\,Y)\,L_a))\,(L'\,X)}{L\,V, \Sigma \longmapsto [V/X]E, \Sigma_2}$$

Finally, finite maps of the form $\mathsf{case}[\overrightarrow{V \mapsto L}]$ on first-order values are used internally by the execution. Application rules are straightforward as shown in rules *Apply-Case-1* and *Apply-Case-2*: if the application has been seen before, we reuse the result address, otherwise we return a fresh address and remember the new result in the table.

$$refine(\Sigma, L, \mathsf{dep?}) = \Sigma[L_x \mapsto \bullet, L_c \mapsto \bullet, L \mapsto (L_x \to \lambda X.(L_c\ X))]$$
$$\text{where } \Sigma(L) = \bullet \text{ and } L_x, L'_c \notin dom(\Sigma)$$
$$refine(\Sigma, L, V_i) = \Sigma[L \mapsto \bullet^{\vec{V} \cup V_i}] \quad \text{where } \Sigma(L) = \bullet^{\vec{V}}$$
$$refine(\Sigma, L, V) = \Sigma \quad \text{otherwise}$$

Fig. 2. Refinement for Symbolic $\lambda_C$

*Apply-Case-1*
$$\frac{\Sigma(L) = \mathsf{case}[\ldots, V' \mapsto L', \ldots] \quad \delta(\Sigma, =, V\ V') \ni \langle \mathsf{1}, \Sigma' \rangle}{L\ V, \Sigma \longmapsto L', \Sigma'}$$

*Apply-Case-2*
$$\Sigma(L) = \mathsf{case}[V' \mapsto L' \ldots]$$
$$\frac{\delta(\Sigma, =, V_n\ V'_i) \ni \langle \mathsf{0}, \Sigma' \rangle \text{ for all } V'_i \in \{V' \ldots\} \quad L_n \notin dom(\Sigma)}{L\ V_n, \Sigma \longmapsto L', \Sigma[L_x \mapsto \bullet, L \mapsto \mathsf{case}[V' \mapsto L' \ldots, V_n \mapsto L_n]]}$$

### 3.3 Primitive operations

Primitive operations are the primary place where unknown values in the heap are refined, in concert with successful contract checks. Figure 3 shows $\delta$'s definition.

The first four rules cover primitive predicate checks. Ambiguity never occurs for concrete values, and an abstract value may definitely prove or refute the predicate if the available information is enough for the conclusion. If the proof system cannot decide a definite result for the predicate check, $\delta$ conservatively includes *both* answers in the possible results and records assumptions chosen for each non-deterministic branch in the appropriate heap. Rules for partial functions such as addition and integer equality, which fail when given non-numeric inputs, reveal possible refinements when applying. This mechanism, when combined with the SMT-aided proof system given below, is sufficient to provide the precision necessary to prove the absence of contract errors.

### 3.4 SMT-aided proof system

Contract checking and primitive operations rely on a proof system to statically relate values and contracts. We write $\Sigma \vdash V : V_c\ \checkmark$ to mean value $V$ satisfies contract $V_c$, where all addresses in $V$ are defined in $\Sigma$. In other words, under any possible instantiation of the unknown values in $\Sigma$, it would satisfy $V_c$ when checked according to the semantics. On the other hand, $\Sigma \vdash V : V_c\ ✗$ indicates that $V$ definitely fails $V_c$. Finally, $\Sigma \vdash V : V_c\ ?$ is a conservative answer when information from the heap and refinement set is insufficient to draw a definite conclusion. The precision of our analysis depends on the precision of this provability relation—increasing the number of contracts that can be related statically to values prunes spurious paths and eliminates impossible error cases.

*Pred-True*
$$\frac{\Sigma \vdash V : O_? \checkmark}{\delta(\Sigma, O_?, V) \ni \langle 1, \Sigma \rangle}$$

*Pred-False*
$$\frac{\Sigma \vdash V : O_? \text{✗}}{\delta(\Sigma, O_?, V) \ni \langle 0, \Sigma \rangle}$$

*Pred-Ambig-True*
$$\frac{\Sigma \vdash V : O_? \text{?}}{\delta(\Sigma, O_?, V) \ni \langle 1, \text{refine}(\Sigma, \text{V}, \text{O}_?) \rangle}$$

*Pred-Ambig-False*
$$\frac{\Sigma \vdash V : O_? \text{?}}{\delta(\Sigma, O_?, V) \ni \langle 0, \text{refine}(\Sigma, \text{V}, \text{O}_?) \rangle}$$

*Plus-Concrete*
$$\overline{\delta(\Sigma, +, n_1, n_2) \ni \langle n_1 + n_2, \Sigma \rangle}$$

*Plus-Error-1*
$$\frac{\delta(\Sigma, \text{int?}, V_1) \ni \langle 0, \Sigma_1 \rangle}{\delta(\Sigma, +, V_1, V_2) \ni \langle \text{blame}_\Lambda, \Sigma_1 \rangle}$$

*Plus-Error-2*
$$\frac{\delta(\Sigma, \text{int?}, V_1) \ni \langle 1, \Sigma_1 \rangle \quad \delta(\Sigma_1, \text{int?}, V_2) \ni \langle 0, \Sigma_2 \rangle}{\delta(\Sigma, +, V_1, V_2) \ni \langle \text{blame}_\Lambda, \Sigma_2 \rangle}$$

*Plus-Abstract*
$$\frac{\delta(\Sigma, \text{int?}, V_1) \ni \langle 1, \Sigma_1 \rangle \quad \delta(\Sigma_1, \text{int?}, V_2) \ni \langle 1, \Sigma_2 \rangle \quad V_1 \neq n \text{ or } V_2 \neq n}{\delta(\Sigma, +, V_1, V_2) \ni \langle \bullet^{\{\text{int?},(=/\text{c } (+ \ V_1 \ V_2))\}}, \Sigma_2 \rangle}$$

*Eq-Error-1*
$$\frac{\delta(\Sigma, \text{int?}, V_1) \ni \langle 0, \Sigma_1 \rangle}{\delta(\Sigma, =, V_1, V_2) \ni \langle \text{blame}_\Lambda, \Sigma_1 \rangle}$$

*Eq-Error-2*
$$\frac{\delta(\Sigma, \text{int?}, V_1) \ni \langle 1, \Sigma_1 \rangle \quad \delta(\Sigma_1, \text{int?}, V_2) \ni \langle 0, \Sigma_2 \rangle}{\delta(\Sigma, =, V_1, V_2) \ni \langle \text{blame}_\Lambda, \Sigma_2 \rangle}$$

*Eq-True*
$$\frac{\delta(\Sigma, \text{int?}, V_1) \ni \langle 1, \Sigma_1 \rangle \quad \delta(\Sigma_1, \text{int?}, V_2) \ni \langle 1, \Sigma_2 \rangle \quad \Sigma_2 \vdash V_1 : (=/\text{c } V_2) \checkmark}{\delta(\Sigma, =, V_1, V_2) \ni \langle 0, \Sigma_2 \rangle}$$

*Eq-False*
$$\frac{\delta(\Sigma, \text{int?}, V_1) \ni \langle 1, \Sigma_1 \rangle \quad \delta(\Sigma_1, \text{int?}, V_2) \ni \langle 1, \Sigma_2 \rangle \quad \Sigma_2 \vdash V_1 : (=/\text{c } V_2) \text{✗}}{\delta(\Sigma, =, V_1, V_2) \ni \langle 0, \Sigma_2 \rangle}$$

*Eq-Ambig-True*
$$\frac{\delta(\Sigma, \text{int?}, V_1) \ni \langle 1, \Sigma_1 \rangle \quad \delta(\Sigma_1, \text{int?}, V_2) \ni \langle 1, \Sigma_2 \rangle \quad \Sigma_2 \vdash V_1 : (=/\text{c } V_2) \text{?}}{\delta(\Sigma, =, V_1, V_2) \ni \langle 0, \text{refine}(\Sigma_2, \text{V}_1, (=/\text{c } V_2)) \rangle}$$

*Eq-Ambig-False*
$$\frac{\delta(\Sigma, \text{int?}, V_1) \ni \langle 1, \Sigma_1 \rangle \quad \delta(\Sigma_1, \text{int?}, V_2) \ni \langle 1, \Sigma_2 \rangle \quad \Sigma_2 \vdash V_1 : (=/\text{c } V_2) \text{?}}{\delta(\Sigma, =, V_1, V_2) \ni \langle 0, \text{refine}(\Sigma_2, \text{V}_1, (\neq/\text{c } V_2)) \rangle}$$

Fig. 3. Basic Operations

### 3.4.1 Basic proof system

A simple proof system (figure 4) can be obtained which returns definite answers for concrete values, uses heap refinements, and handles negation of predicates and disjointness of data types. We abbreviate $\lambda X.(O_? \ X)$ as $O_?$.

Notice that the proof system only needs to handle a small number of well-understood contracts. We rely on evaluation to naturally break down complex contracts into smaller ones and take care of subtle issues such as divergence and crashing. By the time we have $\Sigma(L) = \bullet^{\vec{V}}$, we can assume all contracts in $\vec{V}$ have terminated with success on $L$. With these simple and obvious rules, our system can already verify a significant number of interesting programs. With SMT solver integration, as described below, we can handle far

$$\frac{\Sigma \vdash \Sigma(L) : V_c \checkmark}{\Sigma \vdash L : V_c \checkmark} \qquad \frac{\Sigma \vdash \Sigma(L) : V_c \text{\ding{55}}}{\Sigma \vdash L : V_c \text{\ding{55}}} \qquad \overline{\Sigma \vdash n : \texttt{int?} \checkmark} \qquad \overline{\Sigma \vdash \lambda X.E : \texttt{proc?} \checkmark}$$

$$\frac{}{\Sigma \vdash V \rightarrow \lambda X.C : \texttt{dep?} \checkmark} \qquad \frac{\Sigma \vdash V : \texttt{int?} \checkmark \quad \Sigma \vdash V : (\texttt{=/c}~0) \checkmark}{\Sigma \vdash V : \texttt{zero?} \checkmark}$$

$$\frac{\Sigma \vdash V : \texttt{int?} \checkmark \quad \Sigma \vdash V : (\texttt{=/c}~0) \text{\ding{55}}}{\Sigma \vdash V : \texttt{zero?} \text{\ding{55}}} \qquad \frac{\Sigma \vdash V : \texttt{int?} \text{\ding{55}}}{\Sigma \vdash V : \texttt{zero?} \text{\ding{55}}}$$

$$\frac{\Sigma \vdash V : O'_? \checkmark \quad O'_? \neq O_? \quad O_?, O'_? \in \{\texttt{int?}, \texttt{proc?}, \texttt{dep?}\}}{\Sigma \vdash V : O_? \text{\ding{55}}} \qquad \overline{\Sigma \vdash \bullet^{\{...V_c...\}} : V_c \checkmark}$$

$$\frac{\Sigma \vdash V : \lambda X.E \checkmark}{\Sigma \vdash V : \lambda X.(\texttt{zero?}~E) \text{\ding{55}}} \qquad \frac{\Sigma \vdash V : \lambda X.E \text{\ding{55}}}{\Sigma \vdash V : \lambda X.(\texttt{zero?}~E) \checkmark} \qquad \frac{\Sigma(V) = \{...\lambda X.(\texttt{zero?}~E)...\}}{\Sigma \vdash V : \lambda X.E \text{\ding{55}}}$$

$$\frac{\Sigma \vdash V : V_c \checkmark \text{ is not derivable} \quad \Sigma \vdash V : V_c \text{\ding{55}} \text{ is not derivable}}{\Sigma \vdash V : V_c \textbf{?}}$$

Fig. 4. Basic Proof System

more interesting constraints, including relations between numeric values, without requiring an encoding of the full language.

### 3.4.2 Integrating an SMT solver

We extend the simple provability relation by employing an external solver.

We first define the translation $\{\!\{ \cdot \}\!\}$ from heaps, address-value pairs, and address-contract pairs into formulas in solver $S$:

$$\begin{aligned}
\{\!\{ \overrightarrow{(L,V)} \}\!\} &= (\bigwedge \overrightarrow{\{\!\{ L \mapsto V \}\!\}}) \\
\{\!\{ L \mapsto n \}\!\} &= L = n \\
\{\!\{ L \mapsto \bullet^{\vec{C}} \}\!\} &= \bigwedge \overrightarrow{\{\!\{ L : C \}\!\}} \\
\{\!\{ L_0 : (\texttt{>/c}~V_1) \}\!\} &= L_0 > V_1 \\
\{\!\{ L : (\texttt{=/c}~(\texttt{+}~V_0~V_1)) \}\!\} &= L = V_0 + V_1
\end{aligned}$$

The translation of a heap is the conjunction of all formulas generated from translatable refinements. The function is partial, and there are straightforward rules for translating specific pairs of $(L : C)$ where $C$ are drawn from a small set of simple, well-understood contracts.

This mechanism is enough for the system to verify many interesting programs because the analysis relies on evaluation to break down complex, higher-order predicates. Not having a translation for some contract $C$ only reduces precision and does not affect soundness.

Next, the extension $(\vdash_S)$ is straightforward. The old relation $(\vdash)$ is refined by a solver $S$. Whenever the basic relation proves $\Sigma \vdash L : C\,\textbf{?}$, we call out to the solver to try to either prove or refute the claim:

$$\frac{\{\!\{\Sigma\}\!\} \wedge \neg \{\!\{V : V_c\}\!\} \text{ is unsat}}{\Sigma \vdash_S V : V_c \checkmark} \qquad \frac{\{\!\{\Sigma\}\!\} \wedge \{\!\{V : V_c\}\!\} \text{ is unsat}}{\Sigma \vdash_S V : V_c \text{\ding{55}}}$$

The solver-aided relation uses refinements available on the heap to generate premises $\{\!\{\Sigma\}\!\}$. Unsatisfiability of $\{\!\{\Sigma\}\!\} \wedge \neg\{\!\{V : C\}\!\}$ is equivalent to validity of $\{\!\{\Sigma\}\!\} \Rightarrow \{\!\{V : C\}\!\}$, hence value definitely satisfies contract $C$. Likewise, unsatisfiability of $\{\!\{\Sigma\}\!\} \wedge \{\!\{V : C\}\!\}$ means $V$ definitely refutes $C$. In any other case, we relate the value-contract pair to the conservative answer.

### 3.5 Program evaluation

We give a reachable-states semantics to programs: the initial program $P$ is paired with an initial heap that maps each address in the program to a fully opaque value, and eval produces all states in the reflexive, transitive closure of the single-step reduction relation closed under evaluation contexts.

$$\begin{aligned}
&\texttt{eval} \;: P \to \mathscr{P}(\varsigma) \\
&\texttt{eval}(\overrightarrow{M}E) = \{\varsigma \mid \overrightarrow{M} \vdash (E';E), \Sigma_0 \longmapsto^\star \varsigma\} \\
&\quad \text{where } E' = amb(\{1, \overrightarrow{(L_h\,H)}\}), (\texttt{module}\,H\,V_c\,V) \in \overrightarrow{M} \\
&\quad \text{and } \Sigma_0 = \{L \mapsto \bullet \mid L \text{ appears in } P\} \cup \{L_h \mapsto \bullet\} \\
&\quad \text{and } amb\{E\} = E;\; amb\{E_i, E\ldots\} = \texttt{if}\,L_i\,E_i\,(amb\{E\ldots\}), \text{ for each fresh address } L_i
\end{aligned}$$

Modules with unknown definitions, which we call *opaque*, complicate the definition of eval, since they may contain references to concrete modules. If only the main module is considered, an opaque module might misuse a concrete value in ways not visible to the system. We therefore apply an unknown function to each concrete module before evaluating the main expression.

### 3.6 Soundness of abstract semantics

A program with unknown components is an abstraction of a fully-known program. Thus, the semantics of the abstracted program should approximate the semantics of any such concrete version. In particular, any behavior the concrete program exhibits should also be exhibited by the abstract approximation of that program.

However, we must be precise as to which behaviors are relevant. Suppose we have a single concrete module that links against a single opaque module. The semantics of this program should include all of the possible behaviors, both good and bad, of the known module assuming the opaque module always lives up to its contract. We exclude from consideration behaviors that cause the unknown module to be blamed, since it is of course impossible to verify an unknown program. In other words, we try to verify the parts of

*Higher-order symbolic execution for contract verification and refutation*    23

the program that are known, assuming arbitrary, but correct, behavior for the parts of the program that are unknown.[3]

For this reason, the precise semantic account of blame is crucial. The demonic context can introduce blame of both the known and unknown modules; since we can distinguish these parties, it is easy to ignore blame of the unknown context.

In the remainder of this section, we formally define the approximation relation and show that evaluation preserves this relation, i.e. if program $q$ is an approximation of program $p$ ($p$ is like $q$ but with no unknown), then the evaluation of $q$ is an approximation of the evaluation of $p$.

### 3.6.1 Approximation

We write $\varsigma' \sqsubseteq \varsigma$ to mean "$\varsigma$ approximates $\varsigma'$," or "$\varsigma'$ instantiates $\varsigma$," which intuitively means $\varsigma$ stands for a set of states including $\varsigma'$. For example, $\langle 1, \emptyset \rangle \sqsubseteq \langle L, \{L \mapsto \bullet\} \rangle$. Because we restrict the instantiating side to contain no symbolic value, the heap is irrelevant, we abbreviate $\langle E', \emptyset \rangle \sqsubseteq \langle E, \Sigma \rangle$ as $E' \sqsubseteq \langle E, \Sigma \rangle$ and $\langle E'_1, \emptyset \rangle \longmapsto \langle E'_2, \emptyset \rangle$ as $E'_1 \longmapsto E'_2$.

**Consistent instantiation of symbolic values** In order to enforce that each symbolic value is instantiated by one concrete value, we parameterize the relation with a fully concrete heap indicating the instantiation of each symbolic value. For example, expression (+ 1 2) instantiates $\langle (+_{L_1}L_2), \{L_1 \mapsto \bullet, L_2 \mapsto \bullet\} \rangle$, parameterized by $\{L_1 \mapsto 1, L_2 \mapsto 2\}$. A naive definition of the approximation without this parameter would admit a weaker approximation relation not preserved by reduction, where different sub-expressions instantiate symbolic values differently. For example, in the following, suppose we admitted that $E' \sqsubseteq \langle E, \Sigma \rangle$ by straightforward structural induction without enforcing consistent instantiation of labels (because 0, 1, 2 each refines $\langle L, \{L \mapsto \bullet\} \rangle$ individually), we would need to prove that their next states preserve the relation.

$$\begin{array}{rcl} E' & = & \texttt{(if 0 1 2)} \\ \langle E', \Sigma \rangle & = & \langle \texttt{(if } L \ L \ L\texttt{)}, \{L \mapsto \bullet\} \rangle \end{array}$$

The next abstract state, however, does not continue to approximate the concrete one:

$$\begin{array}{rcl} E' & \longmapsto & 2 \\ \langle E, \{L \mapsto \bullet\} \rangle & \longmapsto & \langle L, \{L \mapsto 0\} \rangle \end{array}$$

With a parameter enforcing consistent instantiation of symbolic values, we prevent this "accidental" approximation to establish. In the above example, since there is no instantiation $\Sigma'$ such that $\Sigma'(L) = 0$ and $\Sigma'(L) = 2$, we cannot derive that $E' \sqsubseteq^{\Sigma'} \langle E, \Sigma \rangle$ in the first place. Instead, in the following example, $E' \sqsubseteq^{\Sigma'} (E, \Sigma)$, where $\Sigma' = \{L_0 \mapsto 0, L_1 \mapsto 1, L_2 \mapsto 2\}$:

$$\begin{array}{rl} E' = & \texttt{(if 0 1 2)} \\ E = & \texttt{(if } L_0 \ L_1 \ L_2\texttt{)} \quad \Sigma = \{L_0 \mapsto \bullet, L_1 \mapsto \bullet, L_2 \mapsto \bullet\} \end{array}$$

---

[3] Equivalently, we can think of the execution as implicitly blaming each unknown component for each possible error with a trivially constructed counterexample.

**Omitting behavior from unknown components** Our soundness result does not consider additional errors that blame unknown modules, and therefore we parameterize the approximation relation $\sqsubseteq_{\vec{M}}^{\Sigma}$ with the module definitions $\vec{M}$ to select the opaque modules. We omit these parameters where they are easily inferred to ease notation.

Figure 5 shows the definition of $\sqsubseteq_{\vec{M}}^{\Sigma}$. Each concrete value is approximated by a symbolic value if the heap gives no restriction on the symbolic value's behavior. Further, if the concrete value is known to satisfy a contract, adding that contract to the abstract value preserves the approximation. We write $\Sigma(U)$ to mean a straightforward instantiation of all symbolic values in $U$ according to heap $\Sigma$. We extend the relation $\sqsubseteq_{\vec{M}}^{\Sigma}$ structurally to evaluation contexts $\mathscr{E}$, point-wise to sequences, and to sets of program states.

**Instantiating unknown components** Finally, we justify our choice of instantiating unknown functions to only one specific shape, and show that it is sufficient to approximate all possible interactions between the known and unknown program components.

*Lemma 1* (*Canonical counterexample*)
If $V$ and $V'$ come from different modules, $(\texttt{module}\,H\,V_c\,V') \in \vec{M}$, and $(V\ V') \longmapsto^\star A$ where $A$ is a value or $\texttt{blame}^H$, there exists $\lambda X.E$ such that $(\lambda X.E\ V') \longmapsto^\star A$ and $\lambda X.E$ conforms to $V_r$ in the following grammar:

$$\begin{array}{rcl} V_r & ::= & \lambda X.(\texttt{if}\,(\texttt{proc?}\,X)\,(\texttt{if}\,n\,((V_r\,(X\,V))\,X)\,V)\,(V_m\,X)) \\ V_m & ::= & \lambda X.(\texttt{if}\,(=X\,n)\,V\,(V_m\,X)) \mid \lambda X.V \end{array}$$

*Proof*
Without loss of generality, assume $V$ is non-recursive, bug-free, and does not introduce divergence of its own. (If $V$ is recursive, we unroll it as many time as needed to reproduce the finite trace when applied to $V'$. Further, $V$'s own bug or potential divergence must not have affected the result of its application to $V'$, so we replace the corresponding source code with trivial expressions.)

If the function body $E$ can be decomposed into an evaluation context and a redex $\mathscr{E}[E']$, without loss of generality, we only consider cases where $E'$ contains $X$ and depends on an actual value of $X$ to reduce. (Otherwise, because $E'$ does not contain divergence or error of its own, we can safely "partially evaluate" $E'$ to eliminate any redundant redex.)

We therefore translate $E$ for the following cases. Translation $\{\!\{E\}\!\}_{\vec{V}}$ behaves identically to $E$ up to the finite value set $\vec{V}$ that the free variable $X$ in $E$ can have. The translation terminates by decreasing on $E$'s size.

- Case $E = \mathscr{E}[\texttt{if}\,X\,E_1\,E_2]$:
  $\{\!\{E\}\!\} = (\texttt{if}\,(\texttt{proc?}\,X)\,E_1'\,(V_m\,X))$, where
  — $(\texttt{if}\,(\texttt{proc?}\,X)\,E_1'\,E_2') = \{\!\{\mathscr{E}[E_1]\}\!\}$
  — $V_m$ is a table approximating $E_2$ over $\vec{V}$ for $X$. (because $E_2$ does not have errors and divergence, and $X$ is known to be numbers, the evaluation of $E_2$ for these particular values of $X$ is guaranteed to terminate.)
- Case $E = \mathscr{E}[X\,V]$:
  $\{\!\{E\}\!\} = (\texttt{if}\,(\texttt{proc?}\,X)\,(((\lambda Z.\lambda X.\{\!\{\mathscr{E}[Z]\}\!\})\,(X\,V))\,X)\,(\lambda X.0\,X))$
- Case $E = \mathscr{E}[\texttt{proc?}\,X]$:
  Case analysis on $\mathscr{E}$

- Case $\mathscr{E} = \mathscr{E}'[\text{if } [\,]\ E_1\ (V_m\ X)]$:
  $\{\{E\}\} = \text{if}\ (\texttt{proc?}\ X)\ E_1'\ E_2'$
  , where $(\text{if}\ (\texttt{proc?}\ X)\ E_1'\ E_1'') = \{\{\mathscr{E}'[E_1]\}\}$ and $V_m$ is an approximation of $E_2$ over $\overrightarrow{V}$ for $X$.
- Case $\mathscr{E} = \mathscr{E}'[O\ V'\ldots\ [\,]\ V'\ldots]$: Because $(\texttt{proc?}\ X)$ only evaluates to either 0 or 1, there are 3 cases:
  - $(O\ V'\ldots\ [\,]\ V'\ldots)$ preserves the truth of $(\texttt{proc?}\ X)$: then
    $$\{\{E\}\} = \{\{\mathscr{E}'[\texttt{proc?}\ X]\}\}$$
  - $(O\ V'\ldots\ [\,]\ V'\ldots)$ negates the truth of $(\texttt{proc?}\ X)$: then
    $$\{\{E\}\} = (\text{if}\ (\texttt{proc?}\ X)\ E_1'\ E_2')$$
    where $(\text{if}\ (\texttt{proc?}\ X)\ E_1'\ E'') = \{\{\mathscr{E}'[\texttt{0}]\}\}$
    and $(\text{if}\ (\texttt{proc?}\ X)\ E'\ E_2') = \{\{\mathscr{E}'[\texttt{1}]\}\}$
  - $(O\ V'\ldots\ [\,]\ V'\ldots)$ has constant truth reguardless of $(\texttt{proc?}\ X)$: then $\{\{E\}\} = \{\{\mathscr{E}'[\texttt{0}]\}\}$ or $\{\{\mathscr{E}'[\texttt{1}]\}\}$ depending on the constant truth.
- Case $\mathscr{E} = [\,]$: $\{\{E\}\} = (\text{if}\ (\texttt{proc?}\ X)\ \texttt{1}\ (\lambda X.\texttt{0}\ X))$
- Case $E = \mathscr{E}[\texttt{int?}\ X]$: Similar to the previous case but with the clauses reversed.
- Case $E = \mathscr{E}[O\ V'\ldots\ X\ V'\ldots]$, where $V' ::= V\ |\ X$:
  Because $E$ is bug-free and divergence-free by assumption, and $X$ is first-order, $\{\{E\}\} = (\text{if}\ (\texttt{proc?}\ X)\ \texttt{0}\ (V_m\ X))$, where $V_m$ is constructed as a table mapping each value $V_i$ in $\overrightarrow{V}$ to the evaluation of $[V_i/X]E$ (which terminates).
- Case $E = X$: $\{\{E\}\} = \text{if}\ (\texttt{proc?}\ X)\ X\ X$
- Case $E = V$: $\{\{E\}\} = \text{if}\ (\texttt{proc?}\ X)\ V\ V$

□

With the definition of approximation in hand, we now state the main soundness lemma for the system, which is the basis for relative completeness of counterexamples (3.7) and soundness of contract verification (3.8).

*Lemma 2* (*Soundness of reduction relation*)

If $E_1' \sqsubseteq_{\overrightarrow{M}}^{\Sigma'} \langle E_1, \Sigma_1 \rangle$ and $E_1' \longmapsto E_2'$, then $\langle E_1, \Sigma_1 \rangle \longmapsto^\star \langle E_2, \Sigma_2 \rangle$ such that $E_2' \sqsubseteq_{\overrightarrow{M}}^{\Sigma''} \langle E_2, \Sigma_2 \rangle$ and $\Sigma''$ is consistent with $\Sigma'$, for some $E_2, \Sigma_2$, and $\Sigma''$.

We defer all proofs to the appendix for space.

### *3.7 Soundness and relative completeness of counterexample generation*

The semantics of $\lambda_C$ accumulates a first-order path-invariant as standard in first-order symbolic execution. In addition to this, it also refines the shape of unknown higher-order values. When an evaluation reaches an error state, we query the SMT solver for a model to all first-order values. Plugging this instantiation of first-order values into the heap directly gives us an instantiation of all originally omitted values that reproduces the error. An unknown higher-order value with no constraint on it can be any function, particularly the identify

$$\frac{\textit{Unknown}}{\Sigma(L) = \bullet \quad \Sigma'(L) = U}{U \sqsubseteq_{\vec{M}}^{\Sigma'} \langle L, \Sigma \rangle} \qquad \frac{\textit{Unknown-Refined}}{U \sqsubseteq_{\vec{M}}^{\Sigma'} \langle L, \Sigma \rangle \quad \Sigma(L) = \bullet^{\overrightarrow{U_c}} \quad \emptyset \vdash U : \Sigma'(U') \checkmark}{U \sqsubseteq_{\vec{M}}^{\Sigma'} \langle L, \Sigma[L \mapsto \bullet^{\overrightarrow{U_c} \cup \{U'\}}] \rangle}$$

$$\frac{\textit{Lambda-Unknown}}{\Sigma(L) = \lambda X.E \quad E' \sqsubseteq_{\vec{M}}^{\Sigma'} \langle E, \Sigma \rangle}{\lambda X.E' \sqsubseteq_{\vec{M}}^{\Sigma'} \langle L, \Sigma \rangle} \qquad \frac{\textit{Blame-Ignored}}{(\texttt{module}\, H\, V_c\, L) \in \vec{M} \quad \text{or} \quad H \in \{\dagger, \Lambda\}}{\texttt{blame}_{H'}^H \sqsubseteq_{\vec{M}}^{\Sigma'} \langle E, \Sigma \rangle}$$

$$\frac{\textit{Num}}{n \sqsubseteq_{\vec{M}}^{\Sigma'} \langle n, \Sigma \rangle} \qquad \frac{\textit{If}}{E' \sqsubseteq_{\vec{M}}^{\Sigma'} \langle E, \Sigma \rangle \quad E'_1 \sqsubseteq_{\vec{M}}^{\Sigma'} \langle E_1, \Sigma \rangle \quad E'_2 \sqsubseteq_{\vec{M}}^{\Sigma'} \langle E_2, \Sigma \rangle}{\texttt{if}\, E'\, E'_1\, E'_2 \sqsubseteq_{\vec{M}}^{\Sigma'} \langle \texttt{if}\, E\, E_1\, E_2, \Sigma \rangle} \qquad \frac{\textit{Lambda}}{E' \sqsubseteq_{\vec{M}}^{\Sigma'} \langle E, \Sigma \rangle}{\lambda X.E' \sqsubseteq_{\vec{M}}^{\Sigma'} \langle \lambda X.E, \Sigma \rangle}$$

$$\frac{\textit{App}}{E'_1 \sqsubseteq_{\vec{M}}^{\Sigma'} \langle E_1, \Sigma \rangle \quad E'_2 \sqsubseteq_{\vec{M}}^{\Sigma'} \langle E_2, \Sigma \rangle}{(E'_1\, E'_2) \sqsubseteq_{\vec{M}}^{\Sigma'} \langle (E_1\, E_2), \Sigma \rangle} \qquad \frac{\textit{Prim}}{E'_i \sqsubseteq_{\vec{M}}^{\Sigma'} \langle E_i, \Sigma \rangle, \text{ for each } E'_i \in \vec{E'}, E_i \in \vec{E}}{(O\, \vec{E'}) \sqsubseteq_{\vec{M}}^{\Sigma'} \langle (O\, \vec{E}), \Sigma \rangle}$$

$$\frac{\textit{Check}}{C' \sqsubseteq_{\vec{M}}^{\Sigma'} \langle C, \Sigma \rangle \quad E' \sqsubseteq_{\vec{M}}^{\Sigma'} \langle E, \Sigma \rangle}{\texttt{mon}_{H''}^{H,H'}(C', E') \sqsubseteq_{\vec{M}}^{\Sigma'} \langle \texttt{mon}_{H''}^{H,H'}(C, E), \Sigma \rangle} \qquad \frac{\textit{Var}}{X \sqsubseteq_{\vec{M}}^{\Sigma'} \langle X, \Sigma \rangle} \qquad \frac{\textit{Blame}}{\texttt{blame}_{H''}^H \sqsubseteq_{\vec{M}}^{\Sigma'} \langle \texttt{blame}_{H''}^H, \Sigma \rangle}$$

$$\frac{\textit{Dep}}{C' \sqsubseteq_{\vec{M}}^{\Sigma'} \langle C, \Sigma \rangle \quad E' \sqsubseteq_{\vec{M}}^{\Sigma'} \langle E, \Sigma \rangle}{C' \rightarrow \lambda X.E' \sqsubseteq_{\vec{M}}^{\Sigma'} \langle C \rightarrow \lambda X.E, \Sigma \rangle} \qquad \frac{}{\Sigma' \sqsubseteq \emptyset} \qquad \frac{\Sigma'_1 \sqsubseteq \Sigma'_2 \quad V' \sqsubseteq^{\Sigma_1} \langle V, \Sigma'_2 \rangle}{\Sigma'_1[L \mapsto V'] \sqsubseteq \Sigma'_2[L \mapsto V]}$$

Fig. 5. Approximation

function that can be simplified away. The remarkable result is that our method of finding counterexamples is both sound and relatively complete with respect to the underlying first-order SMT solver.

**Soundness of counterexamples** Because we refine unknown functions to have specific shapes in addition to maintaining a complete path condition of first-order values, the semantics of $\lambda_C$ is a sound under-approximation of all valid program runs. Therefore, any valid instantiation of the path condition for a specific branch will reproduce the execution following that branch. In particular, an instantiation in an error branch yields a true counterexample triggering the contract violation of that branch.

*Theorem 1* (*Soundness of Counterexamples*)
If $\langle E_1, \Sigma_1 \rangle \longmapsto^\star \langle \texttt{blame}_{H'}^H, \Sigma_n \rangle$, $\Sigma' \sqsubseteq \Sigma_n$, and $E'_1 = \Sigma'(E_1)$ then $E'_1 \longmapsto^\star \texttt{blame}_{H'}^H$.

**Relative completeness of counterexamples** The abstract reduction semantics of $\lambda_C$ also provides a sound over-approximation of all possible interactions between the known and unknown program components, discovering every reachable error in the concrete modules (lemma 2). Therefore, as long as the SMT solver can construct a model to the given first-order formula, we can construct a higher-order function that reproduces each discovered error, simply by plugging in the first-order values given by the solver.

*Theorem 2* (*Relative Completeness of Counterexamples*)
If $E'_1 \longmapsto^\star \mathtt{blame}^H_{H'}$, $E'_1 \sqsubseteq \langle E_1, \Sigma_1 \rangle$, and there is a complete procedure for generating values satisfying first-order constraints, then $\langle E_1, \Sigma_1 \rangle \longmapsto^\star \langle \mathtt{blame}^H_{H'}, \Sigma_n \rangle$ such that we can derive some $\Sigma'$ such that $\Sigma' \sqsubseteq \Sigma_n$.

### 3.8 From bug-finding to verification

The semantics of $\lambda_C$ not only is helpful for generating test cases that reproduce contract violations, it also helps verification of contract-correctness. Because the existence of a counterexample implies the existence of a "canonical" counterexample of the form in rule *Apply-Unknown* (lemma 1), proving the absence of counterexamples of this form alone is equivalent to verification of the program. Unfortunately, a naive run of a program in this semantics does not terminate for most programs: execution unfolds indefinitely to explore an infinite set of instantiations to abstract values. We therefore introduce two transformations that approximate the semantics of $\lambda_C$ to accelerate convergence, making it a practical verification for many programs.

#### 3.8.1 Approximating unknown functions

Rule *Apply-Unknown* shown in section 3.2.3 unfolds and remembers the shape of each unknown function as execution progresses. Although this refinement is useful for constructing higher-order counterexamples, it is a major source of non-termination: the execution repeatedly generates fresh λ-terms. As a more approximate execution of opaque function applications, we no longer unfold an unknown function upon application and replace rule *Apply-Unknown* with two non-deterministic rules: *Apply-Unknown-Success* returns a fresh address approximating an unknown result, and *Apply-Unknown-Havoc* passes the argument to a demonic context whose sole purpose is to find reachable blames in the argument: it repeatedly applies the argument to an unknown value, then places the value back into the unknown context. (Even though $V$ may not be a function, the semantics of blames allows us to ignore the potentially erroneous application, which is the responsibility of the unknown component.)

*Apply-Unknown-Success*
$$\frac{\Sigma(L) = \bullet^{\vec{U}} \qquad \delta(\Sigma, \mathtt{proc?}, L) \ni \langle 1, \Sigma' \rangle}{L\, V, \Sigma \longmapsto L', \Sigma'[L' \mapsto \bullet]}$$

*Apply-Unknown-Havoc*
$$\frac{\Sigma(L) = \bullet^{\vec{U}} \qquad \delta(\Sigma, \mathtt{proc?}, L) \ni \langle 1, \Sigma' \rangle}{L\, V, \Sigma \longmapsto L\, (V\, L'), \Sigma'[L' \mapsto \bullet]}$$

This abstraction does not allow easy construction of concrete counterexamples in case of errors, and may introduce more spurious paths, but does not significantly affect precision in practice. Below is an example where the abstracted semantics steps to a false contract violation, even though the second error is not reachable. The unknown function f either applies or ignores its argument, but the abstraction prevents execution from remembering the choice in a particular branch.

```
(let ([f L])  ;; where {L ↦ •}
  (f (λ (x) (/ 1 0)))
  (f (λ (x) (/ 1 x))))
```

*Lemma 3* (*Soundness of unknown function approximation*)
If $(V_1\ V_2) \sqsubseteq \langle (L\ V), \Sigma \rangle$ and $(V_1\ V_2) \longmapsto E'$ then $\langle (L\ V), \Sigma \rangle \longmapsto^\star \langle E, \Sigma' \rangle$ such that $E' \sqsubseteq \langle E, \Sigma' \rangle$.

### 3.8.2 Summarizing function results

With the abstraction as presented in section 3.8.1, the semantics still does not terminate for many common recursive programs. Consider the following example:

```
(define (fact n)
  (if (= n 0) 1 (* n (fact (- n 1)))))
(fact L_n)
```

Ignoring error cases, it eventually reduces non-deterministically to all of the following:

$$1 \text{ if } L_n \mapsto \texttt{0}$$
$$(\texttt{*}\ L_n\ \texttt{1}) \text{ if } L_n \not\mapsto \texttt{0}, L_{n-1} \mapsto \texttt{0}$$
$$(\texttt{*}\ L_n\ (\texttt{*}\ L_{n-1}\ (\texttt{fact}\ L_{n-1}))) \text{ if } L_n, L_{n-1} \not\mapsto \texttt{0}$$

where $L_{n-1}$ is a fresh address resulting from subtracting $L_n$ by one. The process continues with $L_{n-2}, L_{n-3}$, etc. This behavior from the analysis happens because it attempts to approximate *all* possible concrete substitutions to abstract values. Although fact terminates for all concrete naturals, there are an infinite number of those: $L_n$ can be 0, 1, 2, and so on.

To enforce termination for all programs, we can resort to well-known techniques such as finite state or pushdown abstractions (Van Horn and Might 2012). But often those are overkill at the cost of precision. Consider the following program:

```
(let* ([id (λ (x) x)]
       [y (id 0)]
       [z (id 1)])
  (< y z))
```

where a monovariant flow analysis such as 0CFA (Shivers 1988) thinks y and z can be both 0 and 1, and pushdown analysis thinks y is 0 and z is either 0 or 1. For a concrete, straight-line program, such imprecision seems unsatisfactory. We therefore aim for an analysis that provides exact execution for non-recursive programs and retains enough invariants to verify interesting properties of recursive ones. The analysis quickly terminates for a majority of programming patterns with decent precision, although it is not guaranteed to terminate in the general case—see section 4 for empirical results.

One technical difficulty is that the semantics of contracts prevents us from using a recursive function's contract directly as a loop invariant, because contracts are only boundary-level enforcement. It is unsound to assume returned values of internal calls can be approximated by contracts, as in f below.

```
(f : (and/c int? (≥/c 0)) → (and/c int? (≥/c 0)))
(define (f n)
  (if (= n 0) "" (string-length (f (- n 1)))))
```

If we assume the expression (f (- n 1)) returns a number as specified in the contract, we will conclude f never returns, and is blamed either for violating its own contract by returning a string, or for applying string-length to a number. However, f returns 0 when applied to 1. To soundly and precisely approximate this semantics in the absence of types, we recover data type invariants by execution.

We modify the application rules as follows. At each application, we decide whether execution should step to the function's body or wait for known results from other branches. When an application (f v) reduces to a similar application, we plug in known results instead of executing f's body again, avoiding the infinite loop. Correspondingly, when (f v) returns, we plug the new-found answer into contexts that need the result of (f v). The execution continues until it has a set soundly describing the results of (f v).

To track information about application results and waiting contexts, we augment the execution with two global tables $M$ and $\Xi$ as shown in figure 8. We borrow the choice of metavariable names from work on concrete summaries (Johnson and Van Horn 2014).

A value memo table $M$ maps each application to known results and corresponding refinements. Intuitively, if $M(\Sigma, V_f, V_x) \ni (V, \Sigma')$ then in some execution branch, there is an application $(V_f\ V_x), \Sigma \longmapsto^\star (V, \Sigma')$.

A context memo table $\Xi$ maps each application to contexts waiting for its result. Intuitively, $\Xi(\Sigma, V_f, V_x) \ni \langle F, \Sigma', \mathscr{E}_1, \mathscr{E}_k \rangle$ means during evaluation, some expression

$$\mathscr{E}_1[(\text{rt}_{\langle \Sigma, V_f, V_x \rangle}\ [\mathscr{E}_k[(V_f\ V_z)]])]$$

with heap $\Sigma'$ is paused because applying $(V_f\ V_z)$ under assumptions in $\Sigma'$ is the same as applying $(V_f\ V_x)$ under assumptions in $\Sigma$ up to consistent address renaming specified by function $F$.

To keep track of function applications seen so far, we extend the language with the expression ($\text{rt}_{\langle \Sigma, V, V' \rangle}\ E$), which marks $E$ as being evaluated as the result of applying $V$ to $V'$, but otherwise behaves like $E$. The expression ($\text{blur}_{\langle F, \Sigma, V \rangle}\ E$), whose detailed role is discussed below, approximates $E$ under guidance from a "previous" value $V$.

A state in the approximating semantics with summarization consists of global tables $\Xi$, $M$, and a set $S$ of explored states $\overrightarrow{\varsigma}$.

Reduction now relates tables $\Xi$, $M$, and a set of states $\overrightarrow{\varsigma}$ to new tables $\Xi'$, $M'$ and a new set of states $\overrightarrow{\varsigma}'$. We define a relation $\langle \Xi, M, \varsigma \rangle \longmapsto \langle \Xi, M, \varsigma \rangle$, and then lift this relation point-wise to sets of states. Figure 7 only shows rules that use the global tables or new expression forms.

In the first rule, if an application $((\lambda X.E)\ V)$ is not previously seen, execution proceeds as usual, evaluating expression $E$ with $X$ bound to $V$, but marking this expression using rt.

Second, if a previous application of $((\lambda X.E)\ V_0)$ results in application of the same function to a new argument $V$, we approximate the new argument before continuing. Relation $\approx^F$, straightforwardly defined in figure 9, determines whether two states are equivalent to each other up to renaming $F$. Taking advantage of knowledge of the previous argument, we guess the transition from the $V_0$ to $V$ and heuristically emulate an arbitrary amount of such transformation using the $\oplus$ operator.

Third, when an application results in a similar one, we avoid stepping into the function body and use known results from table $M$ instead. In addition, we refine the current heap to make better use of assumptions about the particular "base case". We also remember the current context as one waiting for the result of such application. To speed up convergence, apart from feeding a new answer $V_a$ to the context, we wrap the entire expression inside (blur$_{\langle F,\Sigma,V\rangle}$ [ ]) to approximate the future result.

The fourth rule in figure 7 shows reduction for returning from an application. Apart from the current context, the value is also returned to any known context waiting on the same application. Besides, the value is also remembered in table $M$. The resumption and refinement are analogous to the previous rule.

Finally, expression (blur$_{\langle F,\Sigma,V_0\rangle}$ $V$) approximates value $V$ under guidance from the previous value $V_0$ and also approximates values on the heap from observation of the previous case. Overall, the approximating operator $\oplus$ occurs in three places: arguments of recursive applications, result of recursive applications, and abstract values on the heap when recursive applications return.

Figure 6 shows an implementation of operator $\oplus$ in an extended language with pairs. The operator approximates the right operand with guidance from the left operand. We also extend the syntax of values to represent inductively defined sets of values. For example, $\mu X.\{\texttt{empty},\langle \bullet^{\texttt{int?}}, !X\rangle\}$ denotes a proper list of integers. We approximate a concrete integer to an abstract one if a previous integer has been seen. (A more sophisticated implementation can use more fine-grained approximations such as positive and negative integers.) Approximation of a pair distributes to each component if the left operand is also a pair. If the left operand is an inductively defined set, the new value is "merged" into the set with appropriate renaming or folding. If the left operand syntactically appears in the right one, we emulate an arbitrary number of transitions from the former to the latter with an appropriate inductive set. As a small precision optimization, we unroll the set once, emulating one or more (instead of zero or more) transitions. Finally, we return the value itself as a safe approximation. Notice that it is unsound to approximate an arbitrary value to $\bullet$. In particular, we cannot approximate a concrete function to $\bullet$, discarding code with potential errors to find.[4] A good implementation of $\oplus$ should allow convergence in common cases. Empirical results for our tool are presented in section 4.

**Soundness of summarization:** A system $\langle \Xi, M, S\rangle$ approximates a concrete state $E$ if we can recover $E$ from the system through approximation rules (figure 10). The first rule states that if any state in $S$ approximates $E$, the system approximates it. The second rule states that if the system knows that an instantiation of $(V\ V_x)$ results in a waiting context $\mathscr{E}'_k$, and $E'$ is reachable from a (possibly different) instantiation of $(V\ V_x)$, then the system also approximates $\mathscr{E}'_k[E']$. Context $\mathscr{E}'_0$ is an irrelevant outermost context waiting for the application's result, and context frames (rt$_{\langle \cdot,\cdot,\cdot\rangle}$ [ ]) mark the application history.

As a consequence, summarization properly handles repetition of waiting contexts, and gives results that approximate any number of recursive applications.

With this definition in hand, we can state the central lemma to establish the soundness of the revised semantics that uses summarization.

*Lemma 4* (*Soundness of summarization*)

---

[4] In an implementation using environment instead of substitution, we can distribute the approximation to each closure's environment's range, obtaining approximations such as an inductive set of closures representing an arbitrary number of wrappings around a function.

$$\text{Values} \quad V ::= \ldots \mid \texttt{empty} \mid \langle V, V \rangle \mid \mu X.\overrightarrow{V} \mid !X$$

$$\begin{aligned}
n_0 \oplus n_1 &= \bullet^{\texttt{int?}} &&, \text{if } n_0 \neq n_1 \\
\langle V_0, V_1 \rangle \oplus \langle V_2, V_3 \rangle &= \langle V_0 \oplus V_2, V_1 \oplus V_3 \rangle \\
\mu X.\overrightarrow{V_0} \oplus \mu Y.\overrightarrow{V_1} &= \mu X.(\overrightarrow{V_0} \oplus \overrightarrow{[!X/!Y]V_1}) \\
\mu X.\overrightarrow{V_0} \oplus V &= \mu X.(\overrightarrow{V_0} \oplus [!X/\mu X.\overrightarrow{V_0}]V) \\
V_0 \oplus V_1 &= [(\mu X.\{V_0, [!X/V_0]V_1\})/V_0]V_1 &&, \text{if } V_0 \in_s V_1 \\
V_0 \oplus V_1 &= V_1 &&, \text{otherwise}
\end{aligned}$$

$$\begin{aligned}
\text{where} \quad & V \in_s V \\
& V \in_s \langle V_0, V_1 \rangle, \text{if } V \in_s V_0 \text{ or } V \in_s V_1
\end{aligned}$$

Fig. 6. Approximation

$$\frac{\mathcal{E} \neq \mathcal{E}_1[(\texttt{rt}_{\langle \Sigma_0, \lambda X.E, V_0 \rangle} \mathcal{E}_k)] \text{ for any } \mathcal{E}_1, \mathcal{E}_k, \Sigma_0, V_0}{\langle \Xi, M, \mathcal{E}[((\lambda X.E) \ V)], \Sigma \rangle \longmapsto \langle \Xi, M, \mathcal{E}[(\texttt{rt}_{\langle \Sigma, \lambda X.E, V \rangle} [V/X]E)], \Sigma \rangle}$$

$$\frac{\mathcal{E} = \mathcal{E}_1[(\texttt{rt}_{\langle \Sigma_0, \lambda X.E, V_0 \rangle} \mathcal{E}_k)] \text{ for some } \mathcal{E}_1, \mathcal{E}_k, \Sigma_0, V_0 \quad \langle \Sigma, V \rangle \not\approx \langle \Sigma_0, V_0 \rangle \quad V_1 = V_0 \oplus V}{\langle \Xi, M, \mathcal{E}[((\lambda X.E) \ V)], \Sigma \rangle \longmapsto \langle \Xi, M, \mathcal{E}[(\texttt{rt}_{\langle \Sigma, \lambda X.E, V_1 \rangle} [V_1/X]E)], \Sigma \rangle}$$

$$\frac{\begin{array}{c}\mathcal{E} = \mathcal{E}_1[(\texttt{rt}_{\langle \Sigma_0, V_f, V_0 \rangle} \mathcal{E}_k)] \text{ for some } \mathcal{E}_1, \mathcal{E}_k, \Sigma_0, V_0 \\ \langle \Sigma, V \rangle \approx^F \langle \Sigma_0, V_0 \rangle \quad \Xi' = \Xi \sqcup [\langle \Sigma_0, V_f, V_0 \rangle \mapsto \langle F, \Sigma, \mathcal{E}_1, \mathcal{E}_k \rangle] \\ \langle V_a, \Sigma_a \rangle \in M[\langle \Sigma_0, V_f, V_0 \rangle] \quad \Sigma' = \Sigma \overrightarrow{[L_n \mapsto \Sigma_a[L_o]]} \text{ where } \overrightarrow{\langle L_o, L_n \rangle} = F\end{array}}{\langle \Xi, M, \mathcal{E}[(V_f \ V)], \Sigma \rangle \longmapsto \langle \Xi', M, \mathcal{E}_1[(\texttt{rt}_{\langle \Sigma_0, V_f, V_0 \rangle} (\texttt{blur}_{\langle F, \Sigma_a, V_a \rangle} \mathcal{E}_k[V_a]))], \Sigma' \rangle}$$

$$\frac{\overrightarrow{\langle L_o, L_n \rangle} = F \quad \Sigma' = \Sigma \overrightarrow{[L_n \mapsto \Sigma_0(L_0) \oplus \Sigma(L_0)]}}{\langle \Xi, M, \mathcal{E}[(\texttt{blur}_{\langle F, \Sigma_0, V_0 \rangle} V)], \Sigma \rangle \longmapsto \langle \Xi, M, \mathcal{E}[V_0 \oplus V], \Sigma' \rangle}$$

$$\frac{M' = M \sqcup [\langle \Sigma_0, V_f, V_0 \rangle \mapsto \langle V, \Sigma \rangle]}{\langle \Xi, M, \mathcal{E}[(\texttt{rt}_{\langle \Sigma_0, V_f, V_0 \rangle} V)], \Sigma \rangle \longmapsto \langle \Xi, M', \mathcal{E}[V], \Sigma \rangle}$$

$$\frac{M' = M \sqcup [\langle \Sigma_0, V_f, V_0 \rangle \mapsto \langle V, \Sigma \rangle]}{\langle F, \Sigma_k, \mathcal{E}_1, \mathcal{E}_k \rangle \in \Xi[\langle \Sigma_0, V_f, V_0 \rangle] \quad \Sigma'_k = \Sigma_k \overrightarrow{[L_n \mapsto \Sigma(L_o)]} \text{ where } \overrightarrow{\langle L_o, L_n \rangle} = F}{\langle \Xi, M, \mathcal{E}[(\texttt{rt}_{\langle \Sigma_0, V_f, V_0 \rangle} V)], \Sigma \rangle \longmapsto \langle \Xi, M', \mathcal{E}_1[(\texttt{rt}_{\langle \Sigma_0, V_f, V_0 \rangle} (\texttt{blur}_{\langle F, \Sigma, V \rangle} \mathcal{E}_k[V]))], \Sigma'_k \rangle}$$

Fig. 7. Summarizing Semantics

If $E'_1 \sqsubseteq \langle \Xi_1, M_1, S_1 \rangle$ and $E'_1 \longmapsto E'_2$ then $\langle \Xi_1, M_1, S_1 \rangle \longmapsto^\star \langle \Xi_2, M_2, S_2 \rangle$ such that $E'_2 \sqsubseteq \langle \Xi_2, M_2, S_2 \rangle$.

The proof is given in the appendix. With this lemma in place, it is straightforward to define verification as a simple corollary of soundness and prove a blame theorem.

First we defined when a module is *verified* by our approach.

*Definition 1* (*Verified module*)

$$\begin{array}{lll}
\text{Expressions} & E \mathrel{+}= (\mathsf{rt}_{\langle \Sigma,V,V \rangle}\ E) \mid (\mathsf{blur}_{\langle F,\Sigma,V \rangle}\ E) \\
\text{Evaluation contexts} & \mathcal{E} \mathrel{+}= (\mathsf{rt}_{\langle \Sigma,V,V \rangle}\ \mathcal{E}) \mid (\mathsf{blur}_{\langle F,\Sigma,V \rangle}\ \mathcal{E}) \\
\text{Context memo tables} & \Xi ::= \overrightarrow{((\Sigma,V,V),\overrightarrow{(F,\Sigma,\mathcal{E},\mathcal{E})})} \\
\text{Value memo tables} & M ::= \overrightarrow{((\Sigma,V,V),\overrightarrow{(V,\Sigma)})} \\
\text{Renamings} & F ::= \overrightarrow{\langle L,L \rangle}
\end{array}$$

Fig. 8. Syntax extensions for approximation

$$\frac{}{\langle n, \Sigma' \rangle \approx^F \langle n, \Sigma \rangle} \qquad \frac{\langle E', \Sigma' \rangle \approx^F \langle E, \Sigma \rangle}{\langle \lambda X.E', \Sigma' \rangle \approx^F \langle \lambda X.E, \Sigma \rangle} \qquad \frac{\langle L', L \rangle \in F \quad \Sigma'(L') = \Sigma(L)}{\langle L', \Sigma' \rangle \approx^F \langle L, \Sigma \rangle}$$

$$\frac{\langle E'_f, \Sigma' \rangle \approx^F \langle E_f, \Sigma \rangle \quad \langle E'_x, \Sigma' \rangle \approx^F \langle E_x, \Sigma \rangle}{\langle (E'_f\ E'_x), \Sigma' \rangle \approx^F \langle (E_f\ E_x), \Sigma \rangle} \qquad \frac{\langle E'_i, \Sigma' \rangle \approx^F \langle E_i, \Sigma \rangle\ \text{for each}\ i}{\langle (O\ \overrightarrow{E'}), \Sigma' \rangle \approx^F \langle (O\ \overrightarrow{E}), \Sigma \rangle}$$

$$\frac{\langle E'_0, \Sigma' \rangle \approx^F \langle E_0, \Sigma \rangle \quad \langle E'_1, \Sigma' \rangle \approx^F \langle E_1, \Sigma \rangle \quad \langle E'_2, \Sigma' \rangle \approx^F \langle E_2, \Sigma \rangle}{\langle \mathsf{if}\ E'_0\ E'_1\ E'_2, \Sigma' \rangle \approx^F \langle \mathsf{if}\ E_0\ E_1\ E_2, \Sigma \rangle}$$

$$\frac{\langle C', \Sigma' \rangle \approx^F \langle C, \Sigma \rangle \quad \langle E', \Sigma' \rangle \approx^F \langle E, \Sigma \rangle}{\langle C' \mapsto \lambda X.E', \Sigma' \rangle \approx^F \langle C \mapsto \lambda X.E, \Sigma \rangle} \qquad \frac{\langle C', \Sigma' \rangle \approx^F \langle C, \Sigma \rangle \quad \langle E', \Sigma' \rangle \approx^F \langle E, \Sigma \rangle}{\langle \mathsf{mon}_{H''}^{H,H'}(C', E'), \Sigma' \rangle \approx^F \langle \mathsf{mon}_{H''}^{H,H'}(C, E), \Sigma \rangle}$$

$$\frac{\langle V', \Sigma' \rangle \approx^F \langle V, \Sigma \rangle \quad \langle V'_c, \Sigma' \rangle \approx^F \langle V_c, \Sigma \rangle}{\langle \mathsf{assume}(V', V'_c), \Sigma' \rangle \approx^F \langle \mathsf{assume}(V, V_c), \Sigma \rangle} \qquad \frac{}{\langle \mathsf{blame}_{H''}^{H}, \Sigma' \rangle \approx^F \langle \mathsf{blame}_{H'}^{H}, \Sigma \rangle}$$

Fig. 9. State equivalence up to renaming

A module $(\mathsf{module}\ H\ V_c\ V) \in P$ is *verified in P* if $V \neq L$ and $\mathit{eval}(P) \not\ni \mathsf{blame}^H$.

Now, by soundness, $H$ is always safe.

*Theorem 3* (*Verified modules can't be blamed*)

If a module named $H$ is verified in $P$, then for any concrete program $Q$ for which $P$ is an abstraction, $\mathit{eval}(Q) \not\ni \mathsf{blame}^H$.

$$\frac{E' \sqsubseteq \langle E, \Sigma \rangle \quad \langle E, \Sigma \rangle \in S}{E' \sqsubseteq \langle \Xi, M, S \rangle}$$

$$\frac{\begin{array}{c}\Xi(\Sigma_0, V, V_x) = \langle F, \Sigma_1, \mathcal{E}_0, \mathcal{E}_k \rangle \\ \mathcal{E}'_0 \sqsubseteq \langle \mathcal{E}_0, \Sigma_1 \rangle \quad \mathcal{E}'_k \sqsubseteq \langle \mathcal{E}_k, \Sigma_1 \rangle \quad V' \sqsubseteq \langle V, \Sigma_1 \rangle \quad V'_0 \sqsubseteq \langle V_x, \Sigma_1 \rangle \quad V'_1 \sqsubseteq \langle V_x, \Sigma_1 \rangle \\ \mathcal{E}'_0 \neq \mathcal{E}'_1[(\mathsf{rt}_{\langle \_,V',\_ \rangle}\ \mathcal{E}'_2)]\ \text{for any}\ \mathcal{E}'_1, \mathcal{E}'_2 \quad \mathcal{E}'_0[(\mathsf{rt}_{\langle \emptyset,V',V'_1 \rangle}\ E')] \sqsubseteq \langle \Xi, M, S \rangle\end{array}}{\mathcal{E}'_0[(\mathsf{rt}_{\langle \emptyset,V',V'_0 \rangle}\ \mathcal{E}'_k[(\mathsf{rt}_{\langle \emptyset,V',V'_1 \rangle}\ E')])] \sqsubseteq \langle \Xi, M, S \rangle}$$

Fig. 10. Approximation of Summarizing Semantics

## 4 Implementation and evaluation

To validate our approach, we implemented a static contract checking tool, SCV, based on the semantics presented in section 3. The system refutes incorrect programs with concrete test cases by running the semantics in section 3.2 and verifies the absence of run-time errors in correct programs using the abstracted semantics in section 3.8.2. In addition, there are a number of implementation extensions for increased precision and performance. We then applied SCV to a wide selection of programs drawn from the literature on verification of higher-order programs, and report on the results.

The source code for SCV and all benchmarks are available along with instructions on reproducing the results we report.[5] Apart from being implemented as a command line tool, our prototype is also available as a public web REPL.[6]

### *4.1 Implementation extensions*

SCV supports an extended language beyond that presented in section 3 in order to handle realistic programs. First, more base values and primitive operations are supported, such as strings and symbols (and their operations), although we do not yet use a solver to reason about values other than integers. We support Racket's numeric tower, which introduces more error sources and interesting counterexamples. Second, data structure definitions are allowed at the top-level. Each new data definition induces a corresponding (automatic) extension to the refinement of unknown functions to deal with the new class of data. The unknown function now also non-deterministically decomposes its argument if the argument is a user-defined struct, in addition to applying functions and mapping first-order values as in rule *Apply-Unknown*. We also extend the widening operator $\oplus$ to heuristically approximate values of user-defined structs to inductively defined data, which gives good precision in common programs. Third, modules have multiple named exports, to handle the examples presented in section 2, and can include local, non-exported, definitions. Fourth, functions can accept multiple arguments and can be defined to have variable-arity, as with `+`, which accepts arbitrarily many arguments. This introduces new possibilities of errors from arity mismatches. Fifth, a much more expressive contract language is implemented with `and/c`, `or/c`, `struct/c`, `µ/c` for conjunctive, disjunctive, data type, and recursive contracts, respectively. Sixth, we provide solver back-ends for both CVC4 (Barrett et al. 2011) and Z3 (Moura and Bjørner 2008).

### *4.2 Evaluating on existing benchmarks*

To evaluate the applicability of SCV to a wide variety of challenging higher-order contract checking problems, we collect examples from the following sources: programs that make use of control-flow-based typing from work on **occurrence typing** (Tobin-Hochstadt and Felleisen 2010), programs from work on **soft typing**, which uses flow analysis to check the preconditions of operations (Cartwright and Fagan 1991), programs with sophisticated specifications from work on model checking **higher-order recursion schemes** (Kobayashi et al. 2011), programs from work on inference of **dependent refinement types** (Terauchi 2010), and programs with rich contracts from our prior work on **higher-order symbolic execution** (Tobin-Hochstadt and Van Horn 2012). We also evaluate SCV on three interactive student video games built for a first-year programming course: **Snake**, **Tetris**, and **Zombie**. These programs were all originally written as sample solutions, following the style expected of students in the course. Of these, Zombie is the most interesting: it was originally an object-oriented program, translated using the encoding seen in section 2.6. Finally, we collect programs submitted anonymously by the users of our web service.

---

[5] `github.com/philnguyen/soft-contract`
[6] `scv.umiacs.umd.edu`

| Corpus | Lines | Checks | Correct Variant (ms) | Incorrect Variant (ms) |
|---|---|---|---|---|
| Occurrence Typing | 116 | 141 | 98.7 | 502.8 |
| Soft Typing | 134 | 177 | 12,747.0 | 331.0 |
| Higher-order Recursion Schemes | 527 | 859 | 14,190.7 (8) | 8,172.3 |
| Dependent Refinement Types | 69 | 116 | 576.7 | 2,270.7 |
| Higher-order Symbolic Execution | 223 | 308 | 9,532.0 (1) | 633.8 |
| Correct anonymous programs (22) | 158 | 213 | 268.6 | - |
| Incorrect anonymous programs (110) | 778 | 1,336 | - | 14,126.9 (5) |
| Student Video Games | | | | |
|   Snake | 164 | 246 | 38,602.3 | 3,034.2 |
|   Tetris | 267 | 338 | 12,303.5 | 2,255.0 |
|   Zombie | 249 | 476 | 21,276.2 | 1,152.0 |

Table 1. Summary benchmark results. (See the appendix for detailed results.)

In order to evaluate our counterexample generation, we modify many correct programs to introduce errors. To do so, we weakened preconditions, (wrongly) strengthened posconditions, or omitted checks before performing partial operations. For example, a resulting program may deconstruct a potentially empty list, compare potentially non-real numbers, or promise strict inequality where equality may happen in an edge case. We believe these changes are representative of common mistakes.

We present our results in summary form in table 1, grouping each of the above sets of benchmark programs; expanded forms of the tables are provided in the appendix. The table shows total line count (excluding blank lines and comments) and the number of static occurrences of contracts and primitives requiring dynamic checks such as function applications and primitive operations. These checks can be eliminated if we can show that they never fail; this has proven to produce significant speedups in practice, even without eliminating more expensive contract checks (Tobin-Hochstadt et al. 2011).

The tables report the time verifying correct programs and refuting their incorrect variants. Execution times are in milliseconds and measured on a Core i7 2.7GHz laptop with 8GB of RAM. When the tool fails to fully verify a program in the "Correct Variant" column, we report the number of false positives next to verification time. Similarly, when the tool fails to generate a concrete counterexample for a program in the "Incorrect Variant" column, we display the number of warnings (without concrete inputs) next to refutation time.

### *4.3 Discussion*

First, SCV works on benchmarks for a range of previous static analyzers, from type systems to model checking to program analysis.

Second, most programs are analyzed in a reasonable amount of time; the longest remaining analysis time is under 60 seconds. This demonstrates that although the termination acceleration method of section 3.8.2 is not fully general, it is effective for many programming patterns. For example, SCV terminates with good precision on `last` from Wright and Cartwright (1997), which hides recursion behind the Y combinator.

Third, across all benchmarks, over 99% (4201/4210) of the contract checks are statically verified, enabling the elimination both of small checks for primitive operations and expensive contracts; see below for timing results. This result emphasizes the value of static contract checking: gaining confidence about correctness from expensive contracts without actually incurring their cost. In practice, problems such as false positives and failure to construct a concrete counterexample do not render the tool useless for the corresponding programs. False positives reduce confidence about the program's correctness and disable contract optimization, but programmers can still run the programs with safety guaranteed by the familiar contract monitoring semantics. On the other hand, even though SCV cannot construct a counterexample for some programs in practice, it always soundly reports potential contract violation. We discuss current difficulties in section 4.5.

Fourth, there are specific examples where our prototype proves to be a good complement to random testing in discovering contract violations. For example, SCV finds a counterexample to the following program quickly and automatically:

```
(define (f n) (/ 1 (- 100 n)))
```

Be default, QuickCheck does not find this error as it only considers integers from `-99` to `99`. Because QuickCheck treats a program as a black box, this conservative choice is reasonable for fear that the integer may be a loop variable causing the test case to run for a long time (Hughes 2015). In contrast, SCV explores the program's semantics symbolically and discovers `100` as a good test case.

Fifth, the resulting higher-order counterexamples suggest that SCV can produce useful feedback. For example, it is easy for programmers to forget that Racket supports the full numeric tower (St-Amour et al. 2012) and that the predicate `number?` accepts complex numbers. In the following program, `argmin`'s contract is in fact too weak to protect the function. SCV proves `argmin` unsafe by applying it to a specific combination of arguments. First, `f` is given a function that produces a non-real number. Second, `xs` is given a list of length 2, which is the minimum length to trigger a use of `<`.

```
(f : (any/c → number?) (and/c pair? list?) → any/c)
(define (argmin f xs)
  (argmin/acc f (car xs) (f (car xs)) (cdr xs)))

(define (argmin/acc f b a xs)
  (cond
   [(null? xs) a]
   [(< b (f (car xs))) (argmin/acc f a b (cdr xs))]
```

```
  [else (argmin/acc f (car xs) (f (car xs)) (cdr xs))]]))

Contract violation: argmin violates contract with <
Value 0+1i violates contract real?
An example that triggers this violation:
    (argmin (λ (x) 0+1i) (list 0 0))
```

Finally, SCV analyzes the functional encoding of object-oriented programs effectively. Zombie is one such example with extensive use of higher-order functions to encode objects and classes, and the tool can reveal errors buried in delayed function calls. We believe this is a promising first step for generating classes and objects as counterexamples. In the example below, we define interface posn/c that accepts two messages x and y, and function first-quadrant? that tests whether a position is in the first quadrant. The counterexample reveals one conforming implementation to interface posn/c that causes error in the module.

```
  (define posn/c
    ([msg : (one-of/c 'x 'y)]
     → (match msg ['x number?] ['y number?])))

  ; posn/c → boolean?
  (define (first-quadrant? p)
    (and (≥ (p 'x) 0) (≥ (p 'y) 0)))

Contract violation: first-quadrant? violates contract with <
Value 0+1i violates contract real?
An example that triggers this violation:
  (first-quadrant? (λ (msg) (case msg [(x) 0+1i] [(y) 0])))
```

Overall, our experiments show that our approach is able to discover and use invariants implied by conditional flows of control and contract checks. Obfuscations such as multiple layers of abstractions or complex chains of aliases do not impact precision (a common shortcoming of flow analysis).

Finally, soft contract verification is more broadly applicable than the systems from which our benchmarks are drawn, which typically are successful only on their own benchmarks. For example, type systems such as occurrence typing (Tobin-Hochstadt and Felleisen 2010) cannot verify any non-trivial contracts, and most soft typing systems do not consider contracts at all. Systems based on higher-order model-checking (Kobayashi et al. 2011), and dependent refinement types (Terauchi 2010) assume a typed language; encoding our programs using large disjoint unions produces unverifiable results.

This broad applicability is why we are not able to directly compare SCV to these other systems across all benchmarks. Instead, the Simple system serves as a benchmark for a system which does not contain our primary contributions.

### *4.4 Contract optimization*

We also report speedup results for the three most complex programs in our evaluation, which are interactive games designed for first-year programming courses (Snake, Tetris, and Zombie). For each, we recorded a trace of input and timer events while playing the game, and then used that trace to re-run the game (omitting all graphical rendering) both with the contracts that we verified, and with the contracts manually removed. Each game was run 100 times in both modes; the total time is presented below.

| Program | Contracts On (ms) | Contracts Off (ms) |
|---|---|---|
| snake | 475,799 | 59 |
| tetris | 1,127,591 | 186 |
| zombie | 12,413 | 1,721 |

The timing results are quite striking—speedup ranges from over 5x to over 5000x. This does not indicate, of course, that speedups of these magnitudes are achievable for real programs. Instead, it shows that programmers avoid the rich contracts we are able to verify, because of their unacceptable performance overhead. Soft contract verification therefore enables programmers to write these specifications without the run-time cost.

The difference in timing between Zombie and the other two games is intriguing because Zombie uses higher-order dependent contracts extensively, along the lines of `vec/c` from section 2.6, which intuitively should be more expensive. An investigation reveals that most of the cost comes from monitoring flat contracts, especially those that apply to data structures. For example, in Snake, disabling `posn/c`, a simple contract that checks for a `posn` struct with two numeric fields, cuts the run-time by a factor of 4. This contract is repeatedly applied to every such object in the game. In contrast, higher-order contracts, as in the object encodings used in Zombie, delay contracts and avoid this repeated checking.

### *4.5 Limitations and Challenges*

We discuss current limitations of our approach and solutions in mitigating them.

First, our approach does not yet give a way to verify deep structural properties expressed as dependent contracts such as "`map` over a list preserves the `length`" or "all elements in the result of `filter` satisfy the predicate", resulting in the false positives seen in table 1. However, it can already be used to verify many interesting programs because often safety questions depend only on knowledge of top-level constructors. Examples of these patterns appear in programs from Kobayashi et al. (2011) for programs such as `reverse` (see also §2.5), `nil`, and `mem`.

Second, the analysis is prone to combinatorial explosion as inherent in symbolic execution. In practice, most conditionals come from case analyses instead of independent alternatives, and we rely on a precise proof system to eliminate spurious paths. In addition, we avoid excessive state explosion as in rules *Apply-Case-1* and *Apply-Case-2* and defer state splitting until neccessary by encoding the constraint of equal inputs implying equal outputs during translation. Finally, modularity mitigates the problem further, as modules tend to be small, and contracts at boundaries help recovering neccessary precision.

Third, the search for counterexamples can be significantly hindered by complex preconditions, where the input is guarded against a deep, inductively defined property. Execution follows different branches before begin able to generate a valid input to continue verifying the module. A naive breadth-first search is bogged down by a large frontier resulting from different attempts to generate input, most of which are eventually found invalid. To mitigate this slow-down, we identify a class of expressions as likely to lead to counterexamples and prioritize their execution. Specifically, an expression whose innermost contract monitoring is of a first-order property on a concrete module is likely to reveal a bug.[7] In contrast, expressions in the middle of input generation do not have this form, because the innermost contract monitoring isi on the opaque input source. Once the system successfully instantiates a concrete input and turns the program into this "suspect" form, it focuses on exploring this branch with that input instead of trying numerous other inputs in parallel. Using this simple heuristic, we are able to cut the execution time of a module violating the "braun-tree" invariant from non-terminating after 1 hour down to 2 seconds.

Finally, there is a mismatch in the data-types between the solver's data-type and Racket's rich numeric tower. In particular, Racket supports mixed arithmetic between different types of numbers up to complex numbers (St-Amour et al. 2012), while Z3's treatment of numbers resembles that from most statically typed languages, and the solver does not perform well in generating models involving a dynamic restriction of a number's type. Below is an example where SCV fails to generate a counterexample:

```
(f : integer? → integer?)
(define (f n) (/ 1 (+ 1 (* n n))))
```

In Racket, division is defined on the full numeric tower, and the result of `(/ 1 (+ 1 (* n n)))` may not be an integer. In the generated query, this result is an unknown number $L$ of type `Real`, and the solver cannot give a model to a constraint set asserting "`(not (is_int L))`". In addition, Racket distinguishes between exact and inexact numbers, where inexact numbers are floating point approximations. Because Z3 does not reason about floating points, we currently do not soundly model inexact arithmetic.

## 5  Related work

In this section, we relate our work to related strands of research: symbolic execution, random-testing, soft-typing, static contract verification, refinement types, and model checking of recursion schemes.

**Symbolic execution:** Symbolic execution is the idea of running programs with abstract inputs. Symbolic execution on first-order programs is mature and has been used to find bugs in real-world programs (Cadar et al. 2006, 2008). Cadar et al. (2006) presents a symbolic execution engine for C that generates counterexamples of the form of mappings from addresses to bit-vectors. Later work extends the technique to generate comprehensive test cases that discover bugs in large programs interacting with the environment (Cadar et al. 2008).

---

[7] In a symbolic program, the monitored value in this position is usually abstract and covers all values the module produces

**Random Testing:** Random testing is a lightweight technique for finding counterexamples to program specifications through randomly generated inputs. QuickCheck for Haskell (Claessen and Hughes 2000) proves the approach highly practical in finding bugs for functional programs. Later works extend random testing to improve code coverage and scale the technique to more language features such as states and class systems. Heidegger and Thiemann (2010) use contracts to guide random testing for Javascript, allowing users to annotate inputs to combine different analyses for increasing the probability of hitting branches with highly constrained preconditions. Klein et al. (2010) also extend random testing to work on higher-order stateful programs, discovering many bugs in object-oriented programs in Racket. Seidel et al. (2015) use refinement types as generators for tests, significantly improving code coverage.

Our approach is a complement to random testing. By combining symbolic execution with an SMT solver, the method takes advantage of conditions generated by ordinary program code and not just user-annotated contracts. In addition, the approach works well with highly constrained preconditions without further help from users. In contrast, random testing systems typically require programmers to implement custom generators (Claessen and Hughes 2000) or require user annotations to incorporate a specific analysis collecting all literals in the program to guide input construction (Heidegger and Thiemann 2010). Type-targeted testing (Seidel et al. 2015) is more lightweight and does not necessitate an extension to the existing semantics, but gives no guarantee about completeness, as inherent in random testing. Even though the tool rules out test cases that fail the pre-conditions, regular code and post-conditions do not help the test generation process. Our system makes use of both contracts and regular code to guide the execution to seek inputs that both satisfy preconditions and fail post-conditions. Exploring possible combination of symbolic execution and random testing for more efficient bug-finding in higher-order programs is our future work.

**Soft typing:** Verifying the preconditions of primitive operations can be seen as a weak form of contract verification and soft typing is a well studied approach to this kind of verification (Cartwright and Felleisen 1996). There are two predominant approaches to soft-typing: one is based on a generalization of Hindley-Milner type inference (Cartwright and Fagan 1991; Wright and Cartwright 1997; Aiken et al. 1994), which views an untyped program as being embedded in a typed one and attempts to safely eliminate coercions (Henglein 1994). The other is founded on set-based abstract interpretation of programs (Flanagan et al. 1996; Flanagan and Felleisen 1999). Both approaches have proved effective for statically checking preconditions of primitive operations, but the approach does not scale to checking pre- and post-conditions of arbitrary contracts. For example, Soft Scheme (Cartwright and Fagan 1991) is not path-sensitive and does not reason about arithmetic, thus it is unable to verify many of the occurrence-typing or higher-order recursion scheme examples considered in the evaluation.

**Contract verification:** Following in the set-based analysis tradition of soft-typing, there has been work extending set-based analysis to languages with contracts (Meunier et al. 2006). This work shares the overarching goal of this paper: to develop a static contract checking approach for components written in untyped languages with contracts. However the work fails to capture the control-flow-based type reasoning essential to analyzing untyped programs and is unsound (as discussed by Tobin-Hochstadt and Van Horn (2012)). Moreover, the set-based formulation is complex and difficult to extend to features considered here.

Our prior work (Tobin-Hochstadt and Van Horn 2012), as discussed in the introduction, also performs soft contract verification, but with far less sophistication and success. As our empirical results show, the contributions of this paper are required to tackle the arithmetic relations, flow-sensitive reasoning, and complex recursion found in our benchmarks.

An alternative approach has been applied to statically checking contracts in Haskell and OCaml (Xu 2012; Xu et al. 2009), which is to inline monitors into a program following a transformation by Findler and Felleisen (2002) and then simplify the program, either using the compiler, or a specialized symbolic engine equipped with an SMT solver. The approach would be applicable to untyped languages except for the final step dubbed *logicization*, a type-based transformation of program expressions into first-order logic (FOL). A related approach used for Haskell is to use a denotational semantics that can be mapped into FOL, which is then model checked (Vytiniotis et al. 2013), but this approach is highly dependent on the type structure of a program. In contrast, our approach does not assume a type system to guide the verification process, and therefore verifies run-time type-safety in addition to richer contracts. Further, these approaches assume a different semantics for contract checking that monitors recursive calls. This allows the use of contracts as inductive hypotheses in recursive calls. In contrast, our approach can naturally take advantage of this stricter semantics of contract checking and type systems, but can also accommodate the more common and flexible checking policy. Additionally, our approach does not rely on type information, the lack of which makes these approaches inapplicable to many of our benchmarks.

Contract verification in the setting of typed, first-order contracts is much more mature. A prominent example is the work on verifying C# contracts as part of the Code Contracts project (Fähndrich and Logozzo 2011).

**Refinement type checking:** Refinement types are an alternative approach to statically verifying pre- and post-conditions in a higher-order functional language. There are several approaches to checking type refinements; one is to restrict the computational power of refinements so that checking is decidable at type-checking time (Freeman and Pfenning 1991); another is to allow unrestricted refinements as in contracts, but to use a solver to attempt to discharge refinements (Knowles and Flanagan 2010; Rondon et al. 2008; Vazou et al. 2013). In the latter approach, when a refinement cannot be discharged, some systems opt to reject the program (Rondon et al. 2008; Vazou et al. 2013), while others such as hybrid type-checking residualize a run-time check to enforce the refinement (Knowles and Flanagan 2010), similar to the way soft-typing residualizes primitive pre-condition checks. Although the end result of our approach closely resembles that of hybrid type checking, we differ in a few important respects. First, we do not rely on an existing type system. Second, the method scales straightforwardly to first-class contracts, whereas existing refinement type systems allow user-defined predicates only for base types and no mechanism for a dynamically computed mix of flat and higher-order specifications. Third, symbolic execution ignores unreachable errors such as those under unreachable lambdas while type checking eagerly checks all code. Finally, handling unknown functions on the semantics side instead of relying on the theory of uninterpreted functions introduces potentially fewer difficulties in scaling to effectful contracts, and allows straightforward generation of higher-order counterexamples.[8]

DJS (Chugh et al. 2012; Chugh et al. 2012) supports expressive refinement specification and verification for stateful JavaScript programs, including sophisticated dependent specifications which SCV cannot verify. However, most dependent properties require heavy annotations. Moreover, `null` inhabits every object type. Thus the approach cannot give the same guarantees about programs such as `reverse` (§2.5) without significantly more annotation burden. Additionally, it relies on whole program annotation, type-checking, and analysis.

**Model checking higher-order recursion schemes:** Much of the recent work on model checking of higher-order programs relies on the decidability of model checking trees generated by higher-order recursion schemes (HORS) (Ong 2006). A HORS is essentially a program in the simply-typed $\lambda$-calculus with recursion and finitely inhabited base types that generates (potentially infinite) trees. Program verification is accomplished by compiling a program to a HORS in which the generated tree represents program event sequences (Kobayashi 2009b; Kobayashi et al. 2010). This method is sound and complete for the simply typed $\lambda$-calculus with recursion and finite base types, but the gap between this language and realistic languages is significant. Subsequently, an untyped variant of HORS has been developed (Tsukada and Kobayashi 2010), which has applications to languages with more advanced type systems, but despite the name it does not lead to a model checking procedure for the untyped $\lambda$-calculus. A subclass of untyped HORS is the class of recursively typed recursion schemes, which has applications to typed object-oriented programs (Kobayashi and Igarashi 2013). In this setting, model checking is undecidable, but relatively complete with a certain recursive intersection type system (anything typable in this system can be verified). To cope with infinite data domains such as integers, counter-example guided abstraction refinement (CEGAR) techniques have been developed (Kobayashi et al. 2011). The complexity of model checking even for the simply typed case is $n$-EXPTIME hard (where $n$ is the rank of the recursion scheme), but progress on decision procedures (Kobayashi and Ong 2009; Kobayashi 2009a) has lead to verification engines that can verify a number of "small but tricky higher-order functional programs in less than a second."

In comparison, the HORS approach can verify some specifications which SCV cannot, but in a simpler (typed) setting, whereas our lightweight method applies to richer languages. Our approach handles untyped higher-order programs with sophisticated language features and infinite data domains. Higher-order program invariants may be stated as behavioral contracts, while the HORS-based systems only support assertions on first order data. Our work is also able to verify programs with unknown external functions, not just unknown integer values, which is important for modular program verification, and we are able to verify many of the small but tricky programs considered in the HORS work.

---

[8] Solvers such as Z3 and CVC4 do not support model generation for higher-order functions

## 6 Conclusions and perspective

We have presented a lightweight method and prototype implememtation for static contract checking using a non-standard reduction semantics that is capable of verifying and falsifying higher-order modular programs with arbitrarily omitted components. Our tool, SCV, scales to realistic language features such as recursive data structures and modular programs, and verifies programs written in the idiomatic style of dynamic languages. The analysis proves the presence and absence of run-time errors without excessive reliance on programmer help. With zero annotation, SCV already helps programmers find unjustified usage of partial functions by showing concrete inputs that trigger those errors. With explicit contracts, programmers can enforce rich specifications to their programs and have the correct ones optimized away without incurring the significant run-time overhead and incorrect ones quickly falsified with concrete test cases.

While in this paper, we have addressed the problem of soft contract verification, the technical tools we have introduced apply beyond this application. For example, a run of SCV can be seen as a modular program analysis—it soundly predicts which functions are called at any call site. Moreover it can be composed with whole-program analysis techniques to derive modular analyses (Van Horn and Might 2010). Adding temporal contracts (Disney et al. 2011) to our system would produce a model checker for higher-order languages. This breadth of application follows directly from the semantics-based nature of our approach.

**Acknowledgments** We thank Carl Friedrich Bolz, Christos Dimoulas, Jeffrey S. Foster, Michael Hicks, J. Ian Johnson, Robby Findler, Lindsey Kuper, Aseem Rastogi, and Matthew Wilson for comments. We thank John Hughes, Suresh Jagannathan, and Phillipe Meunier for detailed discussion of their respective prior work. We thank Casey Klein for help with Redex, Clayton Menzer for help building the web REPL, and Andrew Ruef for help building a reproducible artifact for PLDI. We thank the anonymous reviewers of OOPSLA 2012, ICFP 2014, and PLDI 2015 for their detailed reviews, which helped to improve the presentation and technical content of the paper. We benefited from discussing preliminary results at the "Dagstuhl Seminar on Scripting Languages and Frameworks: Analysis and Verification" and the "NII Workshop on Software Contracts for Communication, Monitoring, and Security." This material is based on research sponsored by the NSF under award 1218390, the NSA under the Science of Security program, and DARPA under the programs Automated Program Analysis for Cybersecurity (FA8750-12-2-0106) and Cleanslate design of Resilient Adaptive Secure Hosts. The U.S. Government is authorized to reproduce and distribute reprints for Governmental purposes notwithstanding any copyright notation thereon.

## A Proofs

This section presents proofs for theorems in the paper. Lemmas 2 and 4 prove theorems 3. Other lemmas support these main ones.

*Theorem 1* (*Soundness of Counterexamples*)
If $\langle E_1, \Sigma_1 \rangle \longmapsto^\star \langle \text{blame}_{H'}^H, \Sigma_n \rangle$, $\Sigma' \sqsubseteq \Sigma_n$, and $E_1' = \Sigma'(E_1)$ then $E_1' \longmapsto^\star \text{blame}_{H'}^H$.

*Proof*
First, $\Sigma' \sqsubseteq \Sigma_i$ for any heap $\Sigma_i$ on the trace $\langle E_1, \Sigma_1 \rangle \longmapsto^\star \langle \text{blame}_{H''}^H, \Sigma_n \rangle$ (by lemma 6). Next, if $\langle E_i, \Sigma_i \rangle \longmapsto \langle E_{i+1}, \Sigma_{i+1} \rangle$, and $E_i' \sqsubseteq \langle E_i, \Sigma' \rangle$, then $E_i' \longmapsto E_{i+1}'$ such that $E_{i+1}' \sqsubseteq \langle E_{i+1}, \Sigma' \rangle$ (by lemma 7). Therefore, any fully concrete instantiation of the final heap leads the program through the same execution trace. □

*Theorem 2* (*Relative Completeness of Counterexamples*)
If $E_1' \longmapsto^\star \text{blame}_{H'}^H$, $E_1' \sqsubseteq \langle E_1, \Sigma_1 \rangle$, and there is a complete procedure for generating values satisfying first-order constraints, then $\langle E_1, \Sigma_1 \rangle \longmapsto^\star \langle \text{blame}_{H'}^H, \Sigma_n \rangle$ such that we can derive some $\Sigma'$ such that $\Sigma' \sqsubseteq \Sigma_n$.

*Proof*

The discovery of the error follows from soundness of reduction relation (lemma 2). We show that the instantiation of the final heap is relatively complete with respect to the underlying solver by induction on the size of $\Sigma_n$.

- If $\Sigma_n = \emptyset$: There is $\Sigma' = \emptyset$ such that $\emptyset \sqsubseteq \emptyset$.
- If $\Sigma_n = \Sigma_{n-1}[L \mapsto V]$: [9] Assume there is $\Sigma'_{n-1}$ such that $\Sigma'_{n-1} \sqsubseteq \Sigma_{n-1}$.
  - If $V = \bullet^{\vec{C}}$: All constraints in $\vec{C}$ are first-order by construction, which we can produce a model for by assumption. (In particular, if $\vec{C}$, any concrete first-order value can instantiate the unknown value).
  - If $V = \lambda X.E$: Then $\Sigma'_n = \Sigma'_{n-1}[L \mapsto \lambda X.E]$. By induction hypothesis, for each address $L$ in $E$, $\Sigma'_{n-1}(L)$ properly instantiates $\Sigma(L)$.
  - If $V = n$: The case is trivial.

□

*Theorem 3* (*Verified modules can't be blamed*)

If a module named $H$ is verified in $P$, then for any concrete program $Q$ for which $P$ is an abstraction, $eval(Q) \not\ni \texttt{blame}^H$.

*Lemma 1* (*Soundness of abstract reduction relation*)
If $E'_1 \sqsubseteq^{\Sigma'_1}_{\vec{M}} \langle E_1, \Sigma_1 \rangle$ and $E'_1 \longmapsto E'_2$, then $\langle E_1, \Sigma_1 \rangle \longmapsto^\star \langle E_2, \Sigma_2 \rangle$ such that $E'_2 \sqsubseteq^{\Sigma'_2}_{\vec{M}} \langle E_2, \Sigma_2 \rangle$ for some $\Sigma'_2 \supseteq \Sigma'_1$.

*Proof*
By case analysis on the derivation of $E'_1 \longmapsto E'_2$ and $E'_1 \sqsubseteq \langle E_1, \Sigma_1 \rangle$.

- Case $E'_1 = (O \vec{V'_1})$, $E_1 = (O \vec{V_1})$ and $E'_2 = A'_1$ because $\delta(\emptyset, O, \vec{V'_1}) \ni \langle A'_1, \emptyset \rangle$:
  By soundness of $\delta$ (lemma 3), $\langle E_1, \Sigma_1 \rangle \longmapsto \langle E_2, \Sigma_2 \rangle \sqsupseteq E'_2$.
- Case $E'_1 = \texttt{if} \, V' E'_2 E'_f$, $E_1 = \texttt{if} \, V E_2 E_f$ and $E'_1 \longmapsto E'_2$ because $\delta(\emptyset, \texttt{zero?}, V') \ni \langle \texttt{0}, \emptyset \rangle$
  By soundness of $\delta$ (lemma 3), $\delta(\Sigma_1, \texttt{zero?}, V) \ni \langle \texttt{0}, \Sigma_2 \rangle$, so $\langle E_1, \Sigma_1 \rangle \longmapsto \langle E_2, \Sigma_2 \rangle \sqsupseteq E'_2$.
  The other case of conditional is similar.
- Case $E'_1 = (\lambda X.E' \, V'_x)$, $E'_2 = [V'_x/X]E'$, $E_1 = (V_f \, V_x)$:
  - Case $V_f = \lambda X.E$: then $\langle E_2, \Sigma_2 \rangle = \langle [V_x/X]E, \Sigma_1 \rangle \sqsupseteq E'_2$.
  - Case $V_f = L$, where $\Sigma_1(L) = \bullet$: then $\Sigma_2 = \Sigma_1[L \mapsto \lambda X.E]$ as in rule *Apply-Unknown*, and $E'$ is of the restricted form approximated by $E$, so $\langle E_2, [V_x/X]E \rangle \sqsupseteq E'_2$.
  - Case $V_f = L$, where $\Sigma_1(L) = \lambda X.E$: then $\langle E_2, \Sigma_2 \rangle = \langle [V_x/X]E, \Sigma_1 \rangle \sqsupseteq E'_2$.
- Case $E'_1 = \texttt{mon}^{H,H'}_{H''}(V'_c, V')$, $E_1 = \texttt{mon}^{H,H'}_{H''}(V_c, V)$:
  - Case $\delta(\emptyset, \texttt{dep?}, V'_c) \ni \langle \texttt{0}, \emptyset \rangle$: By soundness of $\delta$ (lemma 3), $\delta(\Sigma_1, \texttt{dep?}, V_c) \ni \langle \texttt{0}, \Sigma_2 \rangle$. In addition, by soundness of the provability relation (lemma 4), either both $E'_1$ and $E_1$ take shortcuts to the result, or both step to the contract checking form, or $E'_1$ takes shortcuts and $E_1$ steps to the contract-checking form, which by lemma 5 eventually steps to the result approximating $E'_2$.

---

[9] It is straightforward to see that the heap does not contain cycle, by case analysis on the last step of updating the heap in the reduction relation.

Other cases are straightforward. □

*Lemma 2* (*Soundness of summarization*)
The semantics with summarization using tables $\Xi$ and $M$ is sound with respect to an extension to the original semantics without these tables with trivial rules for rt and blur frames:

$$(\mathrm{rt}_{\langle \_,\_,\_\rangle} V) \longmapsto V$$

$$(\mathrm{blur}_{\langle \_,\_,\_\rangle} V) \longmapsto V$$

If $E'_1 \sqsubseteq \langle \Xi_1, M_1, S_1 \rangle$ and $E'_1 \longmapsto E'_2$, then $\langle \Xi_1, M_1, S_1 \rangle \longmapsto^\star \langle \Xi_2, M_2, S_2 \rangle$ such that $E'_2 \sqsubseteq \langle \Xi_2, M_2, S_2 \rangle$.

*Proof*
By induction on the derivation of $E'_1 \sqsubseteq \langle \Xi_1, M_1, S_1 \rangle$ and case analysis on the reduction $E'_1 \longmapsto E'_2$.

- Case $E'_1 \sqsubseteq \langle \Xi_1, M_1, S_1 \rangle$ because $E'_1 \sqsubseteq \langle E_1, \Sigma_1 \rangle$ and $\langle E_1, \Sigma_1 \rangle \in S_1$:
  Case analysis on $E'_1 \longmapsto E'_2$:
  — Sub-case: $E'_1 = \mathscr{E}'[\lambda X.E'\ V']$ , $E'_2 \longmapsto \mathscr{E}'[(\mathrm{rt}_{\langle \emptyset, \lambda X.E', V' \rangle}\ [V'/X]E')]$ , and $E_1 = \mathscr{E}[\lambda X.E\ V]$:
    – If application $(\lambda X.E\ V)$ is new: $\langle E_1, \Sigma_1 \rangle$ $\beta$-reduces to $\langle E_2, \Sigma_2 \rangle$, and
    
    $$\langle \Xi_1, M_1, S_1 \cup \{\langle E_2, \Sigma_2 \rangle\} \rangle$$
    
    straightforwardly approximates $E'_2$.
    – If application $(\lambda X.E\ V)$ is a recursive call with a new argument: $\langle \Xi_1, M_1, \varsigma_1 \rangle$ $\beta$-reduces with a widened argument, which also straightforwardly approximates $E'_2$.
    – If application $(\lambda X.E\ V)$ is a repeated recursive call:
    
    $$\langle \Xi_1, M_1, S_1 \rangle \longmapsto \langle \Xi_2, M_1, S_2 \rangle$$
    
    where $\Xi_2 = \Xi_1 \sqcup [\langle \Sigma_0, \lambda X.E, V_0 \rangle \mapsto \langle F, \Sigma_1, \mathscr{E}_0, \mathscr{E}_k \rangle]$, and some $S_2 \supseteq S_1$.
    Moreover, we have $\mathscr{E} = \mathscr{E}_0[(\mathrm{rt}_{\langle \Sigma_0, \lambda X.E, V_0 \rangle}\ \mathscr{E}_k)]$, and
    
    $$\mathscr{E}' = \mathscr{E}'_0[(\mathrm{rt}_{\langle \emptyset, \lambda X.E', V'_0 \rangle}\ \mathscr{E}'_k)]$$
    
    so $E'_2 = \mathscr{E}'_0[(\mathrm{rt}_{\langle \emptyset, \lambda X.E', V'_0 \rangle}\ \mathscr{E}'_k[(\mathrm{rt}_{\langle \emptyset, \lambda X.E', V' \rangle}\ [V'/X]E')])]$.
    Because $\langle \mathscr{E}_0[(\mathrm{rt}_{\langle \Sigma_0, \lambda X.E, V_0 \rangle}\ \mathscr{E}_k[\lambda X.E\ V])], \Sigma_1 \rangle \in S_2$, it follows from lemma 9 that
    
    $$\mathscr{E}_0[(\mathrm{rt}_{\langle \Sigma_0, \lambda X.E, V_0 \rangle}\ [V_0/X]E)] \in S_2.$$
    
    Thus, $\mathscr{E}'_0[(\mathrm{rt}_{\langle \emptyset, \lambda X.E', V' \rangle}\ [V'/X]E')] \sqsubseteq \langle \Xi_2, M_1, S_2 \rangle$.
    Hence, $\mathscr{E}'_0[(\mathrm{rt}_{\langle \emptyset, \lambda X.E', V'_0 \rangle}\ \mathscr{E}'_k[(\mathrm{rt}_{\langle \emptyset, \lambda X.E', V' \rangle}\ [V'/X]E')])] \sqsubseteq \langle \Xi_2, M_1, S_2 \rangle$.
  — Other sub-cases are straightforward
- Case $E'_1 \sqsubseteq (\Xi_1, M_1, S_1)$ because:
  — $E'_1 = \mathscr{E}'_0[(\mathrm{rt}_{\langle \emptyset, V', V'_0 \rangle}\ \mathscr{E}'_k[(\mathrm{rt}_{\langle \emptyset, V', V_1 \rangle}\ E')])]$
  — $\mathscr{E}_0[(\mathrm{rt}_{\langle \emptyset, V', V'_1 \rangle}\ E')] \sqsubseteq \langle \Xi_1, M_1, S_1 \rangle$

— $\Xi_1(\Sigma_0, V, V_x) \ni \langle F, \Sigma_1, \mathcal{E}_0, \mathcal{E}_k \rangle$
— $V' \sqsubseteq \langle V, \Sigma_1 \rangle$
— $V'_0 \sqsubseteq \langle V_x, \Sigma_1 \rangle$; $V'_1 \sqsubseteq \langle V_x, \Sigma_1 \rangle$
— $\mathcal{E}'_0 \sqsubseteq \langle \mathcal{E}_0, \Sigma_1 \rangle$; $\mathcal{E}'_k \sqsubseteq \langle \mathcal{E}_k, \Sigma_1 \rangle$

There are 2 subcases, whether $E'$ is a value or can be decomposed into a context and redex.

— If $E'$ is a value $V'_a$:
  This means
  $$E'_1 = \mathcal{E}'_0[(\mathsf{rt}_{\langle \emptyset, V', V'_0 \rangle} \, \mathcal{E}'_k[(\mathsf{rt}_{\langle \emptyset, V', V'_1 \rangle} \, V'_a)])]$$
  $$E'_2 = \mathcal{E}'_0[(\mathsf{rt}_{\langle \emptyset, V', V_0 \rangle} \, \mathcal{E}'_k[V'_a])].$$
  By lemma 10, there exists $\langle \mathcal{E}_0[(\mathsf{rt}_{\langle \Sigma_0, V, V_x \rangle} \, V_a)], \Sigma_1 \rangle \in S_1$ such that
  $$\mathcal{E}'_0[(\mathsf{rt}_{\langle \emptyset, V', V'_0 \rangle} \, V'_a)] \sqsubseteq \langle \mathcal{E}_0[(\mathsf{rt}_{\langle \Sigma_0, V, V_x \rangle} \, V_a)], \Sigma_1 \rangle.$$
  Then $\langle \Xi_1, M_1, S_1 \rangle \longmapsto \langle \Xi_2, M_2, S_2 \rangle$ such that $S_2 \ni \langle \mathcal{E}_0[(\mathsf{rt}_{\langle \Sigma_0, V, V_x \rangle} \, \mathcal{E}_k[V_a])], \Sigma_1 \rangle$, which approximates $E'_2$.

— If $E'_1 = \mathcal{E}'_1[E''_1]$:
  We have $\mathcal{E}'_0[(\mathsf{rt}_{\langle \emptyset, V', V'_1 \rangle} \, \mathcal{E}'_1[E''_1])] \longmapsto \mathcal{E}'_0[(\mathsf{rt}_{\langle \emptyset, V', V'_1 \rangle} \, \mathcal{E}_1[E''_2])]$.
  By induction hypothesis, $\langle \Xi_1, M_1, S_1 \rangle \longmapsto^\star \langle \Xi_2, M_2, S_2 \rangle$, such that
  $$\mathcal{E}'_0[(\mathsf{rt}_{\langle \emptyset, V', V'_1 \rangle} \, \mathcal{E}'_1[E''_2])] \sqsubseteq \langle \Xi_2, M_2, S_2 \rangle.$$
  Because $\Xi_2 \supseteq \Xi_1$, $\mathcal{E}'_0[(\mathsf{rt}_{\langle \emptyset, V', V_0 \rangle} \, \mathcal{E}'_k \mathcal{E}'_1[E''_2])] \sqsubseteq \langle \Xi_2, M_2, S_2 \rangle$ follows.

□

*Lemma 3* (*Soundness of primitive operations*)
If $E' \sqsubseteq^{\Sigma'_1} \langle E, \Sigma_1 \rangle$, $\overrightarrow{V'} \sqsubseteq^{\Sigma'_1} \langle \overrightarrow{V}, \Sigma_1 \rangle$ and $\delta(\emptyset, O, \overrightarrow{V'}) \ni \langle A', \emptyset \rangle$ then $\delta(\Sigma_1, O, \overrightarrow{V}) \ni \langle A, \Sigma_2 \rangle$ such that $A \sqsubseteq^{\Sigma'_2} \langle A, \Sigma_2 \rangle$ and $E' \sqsubseteq^{\Sigma'_2} \langle E, \Sigma_2 \rangle$ for some $\Sigma'_2 \supseteq \Sigma_1$.

*Proof*
By inspection of cases of $O$ and $\overrightarrow{V'} \sqsubseteq \langle \overrightarrow{V}, \Sigma_1 \rangle$ and consistency of the provability relation (lemma 4). □

*Lemma 4* (*Consistency of provability relation*)
If $V' \sqsubseteq \langle V, \Sigma \rangle$ and $V'_c \sqsubseteq \langle V_c, \Sigma \rangle$ then:

- If $\emptyset \vdash V' : V'_c \checkmark$ then either $\Sigma \vdash V : V_c \checkmark$ or $\Sigma \vdash V : V_c\,?$
- If $\emptyset \vdash V' : V'_c \,\chi$ then either $\Sigma \vdash V : V_c \,\chi$ or $\Sigma \vdash V : V_c\,?$
- If $\emptyset \vdash V' : V'_c\,?$ then $\Sigma \vdash V : V_c\,?$

*Proof*
By inspection of cases of $V' \sqsubseteq \langle V, \Sigma \rangle$ and $V'_c \sqsubseteq \langle V_c, \Sigma \rangle$. □

*Lemma 5* (*Soundness of provability relation*)
If $V' \sqsubseteq \langle V, \Sigma_1 \rangle, V'_c \sqsubseteq \langle V_c, \Sigma_1 \rangle, \emptyset \vdash V' : V'_c \checkmark$ and $\Sigma_1 \vdash V : V_c\,?$ then $\langle (V_c \, V), \Sigma_1 \rangle \longmapsto^\star \langle V_a, \Sigma_2 \rangle$ such that $\delta(\Sigma_2, \mathsf{zero?}, V_a) \ni \langle 0, \Sigma_3 \rangle$.

*Proof*

By cases of $\emptyset \vdash V' : V'_c$ ✓ (where $V'$ is concrete) and $V'_c \sqsubseteq \langle V_c, \Sigma_1 \rangle$. □

*Lemma 6*
If $\Sigma' \sqsubseteq \Sigma_2$ and $\langle E_1, \Sigma_1 \rangle \longmapsto \langle E_2, \Sigma_2 \rangle$ then $\Sigma' \sqsubseteq \Sigma_1$.

*Proof*
By case analysis of $\langle E_1, \Sigma_1 \rangle \longmapsto \langle E_2, \Sigma_2 \rangle$. The prior heap is either a restriction of $\Sigma_2$, or has the same domain, mapping some addresses to more abstract values than $\Sigma_2$. □

*Lemma 7* (*Completeness of refinement*)
If $\langle E_1, \Sigma_1 \rangle \longmapsto \langle E_2, \Sigma_2 \rangle$, $\Sigma' \sqsubseteq \Sigma_1$, $\Sigma' \sqsubseteq \Sigma_2$, and $E'_1 \sqsubseteq \langle E_1, \Sigma' \rangle$, then $E'_1 \longmapsto E'_2$ such that $E'_2 \sqsubseteq \langle E_2, \Sigma' \rangle$.

*Proof*
By case analysis on the reduction step. For each case, the reduction leaves enough refinement on the heap to steer all instantiations to the same path. The case on primitve operations is deferred to lemma 8 □

*Lemma 8* (*Completeness of primitive operations*)
If $\delta(\Sigma_1, O, \vec{V}) \ni \langle A, \Sigma_2 \rangle$, $\Sigma' \sqsubseteq \Sigma_1$, $\Sigma' \sqsubseteq \Sigma_2$, and $\vec{V'} \sqsubseteq \langle \vec{V}, \Sigma' \rangle$, then $\delta(\emptyset, O, \vec{V'}) \ni \langle A', \emptyset \rangle$ such that $A' \sqsubseteq \langle A, \Sigma_2 \rangle$.

*Proof*
By inspection of cases of $\delta$. □

*Lemma 9*

If $\langle \emptyset, \emptyset, \{\langle E, \emptyset \rangle\} \rangle \longmapsto^* \langle \Xi, M, S \rangle$ and $\mathscr{E}[(\mathsf{rt}_{\langle \Sigma, V_f, V_x \rangle} E')] \in S$, then $\mathscr{E}[(V_f\ V_x)] \in S$.

*Proof*
By induction on $\langle \emptyset, \emptyset, \{\langle E, \emptyset \rangle\} \rangle \longmapsto^* \langle \Xi, M, S \rangle$.

- Case $\langle \emptyset, \emptyset, \{\langle E, \emptyset \rangle\} \rangle = \langle \Xi, M, S \rangle$: We assume programmers cannot write expressions of the form $(\mathsf{rt}_{\langle \Sigma, V, V \rangle} E)$. The case holds trivially.
- Case $\langle \emptyset, \emptyset, \{\langle E, \emptyset \rangle\} \rangle \longmapsto^* \langle \Xi', M', S' \rangle$ and $\langle \Xi', M', S' \rangle \longmapsto \langle \Xi, M, S \rangle$: Case analysis on the reduction $\langle \Xi', M', S' \rangle$. If the reduction introduces a new frame $(\mathsf{rt}_{\langle \Sigma, V_f, V_x \rangle} E)$ in $S$, it must have resulted from the application $(V_f\ V_x)$ in $S'$.

□

*Lemma 10*

If $\langle \mathscr{E}[(\mathsf{rt}_{\langle \Sigma_0, V_f, V_x \rangle} V)], \Sigma \rangle \sqsubseteq \langle \Xi, M, S \rangle$ where $\mathscr{E} \neq \mathscr{E}_1[(\mathsf{rt}_{\langle \_, V_f, \_ \rangle} \mathscr{E}_2)]$ for any $\mathscr{E}_1, \mathscr{E}_2$, then there is $\varsigma \in S$ such that $\langle \mathscr{E}[(\mathsf{rt}_{\langle \Sigma_0, V_f, V_x \rangle} V)], \Sigma \rangle \sqsubseteq \varsigma$

*Proof*
By case analysis on the derivation

$$\langle \mathscr{E}[(\mathsf{rt}_{\langle \Sigma_0, V_f, V_x \rangle} V)], \Sigma \rangle \sqsubseteq \langle \Xi, M, S \rangle$$

(only the base case of $\cdot \sqsubseteq \langle \cdot, \cdot, \cdot \rangle$ is applicable). □

| | Program | Lines | Checks | Correct Variant (ms) | Incorrect Variant (ms) |
|---|---|---:|---:|---:|---:|
| | ex-01 | 6 | 4 | 3.3 | 32.4 |
| | ex-02 | 6 | 8 | 3.9 | 29.4 |
| | ex-03 | 10 | 12 | 22.0 | 57.8 |
| | ex-04 | 11 | 12 | 7.8 | 41.4 |
| | ex-05 | 6 | 6 | 4.7 | 31.4 |
| Occurrence Typing Examples | ex-06 | 8 | 11 | 5.1 | 32.5 |
| | ex-07 | 8 | 7 | 4.7 | 34.5 |
| | ex-08 | 6 | 11 | 7.0 | 47.2 |
| | ex-09 | 14 | 12 | 8.6 | 32.1 |
| | ex-10 | 6 | 8 | 3.5 | 30.5 |
| | ex-11 | 9 | 8 | 6.7 | 33.3 |
| | ex-12 | 5 | 11 | 5.7 | 31.3 |
| | ex-13 | 9 | 11 | 7.5 | 34.6 |
| | ex-14 | 12 | 20 | 8.2 | 34.4 |
| | **Total** | 116 | 141 | 98.7 | 502.8 |

Table B 1. Logical types for untyped languages benchmarks

| | Program | Lines | Checks | Correct Variant (ms) | Incorrect Variant (ms) |
|---|---|---:|---:|---:|---:|
| | append | 8 | 15 | 22.7 | 6.4 |
| | cpstak | 23 | 15 | 12,449.6 | 46.0 |
| | flatten | 12 | 24 | 27.2 | 37.5 |
| | last-pair | 7 | 9 | 21.1 | 30.6 |
| | last | 17 | 21 | 35.7 | 19.0 |
| Soft Typing Examples | length-acc | 10 | 14 | 26.6 | 8.0 |
| | length | 8 | 13 | 22.7 | 6.7 |
| | member | 8 | 15 | 23.3 | 34.9 |
| | rec-div2 | 9 | 17 | 22.5 | 36.1 |
| | subst* | 11 | 12 | 23.1 | 34.1 |
| | tak | 12 | 14 | 22.7 | 36.8 |
| | taut | 9 | 8 | 22.2 | 34.9 |
| | **Total** | 134 | 177 | 12,719.4 | 331.0 |

Table B 2. Soft typing benchmarks

## B Detailed evaluation results

This section shows detailed evaluation results for benchmarks collected from different verification papers. All are done on a Core i7 @ 2.70GHz laptop running Ubuntu 13.10 64bit. Analysis times are averaged over 10 runs.

| | Program | Lines | Checks | Correct Variant (ms) | Incorrect Variant (ms) |
|---|---|---|---|---|---|
| Higher-order Recursion Scheme Examples | intro1 | 13 | 11 | 26.6 | 208.5 |
| | intro2 | 13 | 11 | 27.7 | 210.2 |
| | intro3 | 13 | 12 | 30.6 | 48.0 |
| | sum | 9 | 12 | 100.4 | 200.4 |
| | mult | 9 | 20 | 188.6 | 221.2 |
| | max | 14 | 11 | 35.7 | 220.2 |
| | mc91 | 8 | 15 | 169.6 (1) | 115.5 |
| | ack | 9 | 16 | 15.8 | 205.5 |
| | repeat | 11 | 11 | 10.1 | 39.7 |
| | fhnhn | 18 | 15 | 38.6 | 64.4 |
| | fold-div | 18 | 34 | 289.0 | 250.2 |
| | hrec | 9 | 13 | 21.8 | 214.1 |
| | neg | 20 | 15 | 95.4 | 255.4 |
| | l-zipmap | 16 | 31 | 483.0 (1) | 152.9 |
| | hors | 25 | 17 | 56.8 | 58.4 |
| | r-lock | 17 | 19 | 75.3 | 90.1 |
| | r-file | 50 | 62 | 84.3 | 118.7 |
| | reverse | 11 | 28 | 20.6 | 288.5 |
| | isnil | 9 | 17 | 14.9 | 9.6 |
| | mem | 12 | 28 | 28.2 | 545.0 |
| | nth0 | 15 | 27 | 24.2 | 806.6 |
| | zip | 14 | 42 | 268.1 | 688.2 |
| | a-max | 18 | 33 | 528 | 294.1 |
| | fold-fun-list | 20 | 32 | 70.5 | 543.2 |
| | fold-left | 14 | 27 | 2028.4 (1) | 145.5 |
| | fold-right | 14 | 27 | 2468.5 (1) | 184.1 |
| | forall-leq | 13 | 23 | 21 | 335 |
| | harmonic | 14 | 26 | 381.6 | 101.2 |
| | length | 13 | 24 | 14.5 (1) | 104.4 |
| | map-filter | 21 | 51 | 2083.1 (1) | 399.3 |
| | risers | 26 | 61 | 38.9 | 267.3 |
| | search | 14 | 26 | 2386.5 (1) | 28.1 |
| | zip-unzip | 27 | 62 | 2064.4 (1) | 758.8 |
| | **Total** | 527 | 859 | 14,190.7 (8) | 8,172.3 |

Table B 3. Higher-order model checking benchmarks

| | Program | Lines | Checks | Correct Variant | Incorrect Variant |
|---|---|---:|---:|---:|---:|
| Dep. Type Inf. | boolflip | 10 | 17 | 10.5 | 38.8 |
| | mult-all | 10 | 18 | 9.2 | 532.8 |
| | mult-cps | 12 | 20 | 348.1 | 52.3 |
| | mult | 10 | 17 | 102.9 | 36.9 |
| | sum-acm | 10 | 15 | 41.1 | 1,132.3 |
| | sum-all | 9 | 15 | 8.9 | 442.1 |
| | sum | 8 | 14 | 9.0 | 35.5 |
| | **Total** | 69 | 116 | 529.7 | 2,270.7 |

Table B 4. Dependent type checking benchmarks

| | Program | Lines | Checks | Correct Variant | Incorrect Variant |
|---|---|---:|---:|---:|---:|
| Symbolic Execution Examples | all | 9 | 16 | 23.0 | 23.2 |
| | even-odd | 10 | 11 | 102.7 | 20.3 |
| | factorial-acc | 10 | 9 | 16.3 | 7.0 |
| | factorial | 7 | 8 | 13.1 | 5.9 |
| | fibonacci | 7 | 11 | 1,345.7 | 97.3 |
| | filter-sat-all | 11 | 18 | 2,053.3 (1) | 23.1 |
| | filter | 11 | 17 | 24.5 | 37.6 |
| | foldl1 | 9 | 17 | 22.0 | 22.2 |
| | foldl | 8 | 10 | 22.6 | 22.4 |
| | foldr1 | 9 | 11 | 22.3 | 21.1 |
| | foldr | 8 | 10 | 27.0 | 23.1 |
| | ho-opaque | 10 | 14 | 19.1 | 19.9 |
| | id-dependent | 8 | 3 | 4.5 | 17.6 |
| | insertion-sort | 14 | 30 | 57.6 | 54.9 |
| | map-foldr | 11 | 20 | 24.2 | 24 |
| | mappend | 11 | 31 | 29.7 | 26.9 |
| | map | 10 | 13 | 23.9 | 23.7 |
| | recip-contract | 7 | 9 | 4.4 | 4.1 |
| | recip | 8 | 15 | 5.7 | 5.5 |
| | rsa | 14 | 5 | 17.7 | 25.1 |
| | sat-7 | 20 | 12 | 5647.7 | 101.8 |
| | sum-filter | 11 | 18 | 25.0 | 27.1 |
| | web (22) | 158 | 213 | 268.6 | - |
| | web (110) | 778 | 1,336 | - | 14,126.9 (5) |
| | **Total** | 1,159 | 1,857 | 9800.6 (1) | 14,760.7 (5) |
| Games | snake | 164 | 246 | 38,602.3 | 3,034.2 |
| | tetris | 267 | 338 | 12,303.5 | 2,255.0 |
| | zombie | 249 | 476 | 21,276.2 | 1,152.0 |
| | **Total** | 680 | 1,060 | 72,182.0 | 6,441.2 |

Table B 5. Higher-order symbolic execution benchmarks